\documentclass[pdftex,aps,english,prb,amsmath,amsfonts,amssymb,superscriptaddress,twocolumn,showpacs,citeautoscript]{revtex4-1}
\usepackage[T1]{fontenc}
\usepackage[latin9]{inputenc}
\usepackage{amsmath}
\usepackage{amssymb}
\usepackage{wasysym}
\usepackage{esint}
\usepackage[pdftex]{graphicx}
\usepackage{times}
\usepackage{nicefrac}
\usepackage{appendix}
\usepackage{textcase}
\usepackage{subfigure}
\usepackage{empheq} 
\usepackage{color}

\makeatletter
\@ifundefined{textcolor}{}
{%
 \definecolor{BLACK}{gray}{0}
 \definecolor{WHITE}{gray}{1}
 \definecolor{RED}{rgb}{1,0,0}
 \definecolor{GREEN}{rgb}{0,1,0}
 \definecolor{BLUE}{rgb}{0,0,1}
 \definecolor{CYAN}{cmyk}{1,0,0,0}
 \definecolor{MAGENTA}{cmyk}{0,1,0,0}
 \definecolor{YELLOW}{cmyk}{0,0,1,0}
}

\renewcommand{\vec}[1]{\mathbf{#1}}
\renewcommand{\Re}{\operatorname{Re}}
\renewcommand{\Im}{\operatorname{Im}}
\newcommand{\tr}{\operatorname{Tr}}
\newcommand{\sym}{\operatorname{sym}}
\newcommand{\h}{\mathcal{H}}

\renewcommand{\o}[1]{{#1}}
\newcommand{\op}[1]{{\o{#1}+}}
\newcommand{\om}[1]{{\o{#1}-}}
\newcommand{\os}[1]{{\o{#1}*}}
\newcommand{\f}{\phi}
\newcommand{\x}{\xi}
\newcommand{\s}{\sigma}
\renewcommand{\b}{\beta}
\newcommand{\G}{\Gamma}
\newcommand{\GD}{\mathbb{G}}
\newcommand{\CD}{\mathbb{C}}
\renewcommand{\S}{\mathcal{S}}
\newcommand{\M}{\mathcal{M}}
\newcommand{\so}{\partial V}
\newcommand{\T}{\mathrm{T}}

\newcommand{\add}[1]{\if\a\b{{\color{red} #1}}\else{#1}\fi}

\newcommand{\xy}{}

\newcommand{\vechat}[1]{\hat{\vec{#1}}}

\newcommand{\pol}{\mathrm{p}}

\newcommand{\citeasnoun}[1]{Ref.~\onlinecite{#1}}

\newcommand{\mat}[2][ccccccccccccccccccc]{
  \left(
  \begin{array}{#1}
    #2
  \end{array}
  \right)}

\renewcommand{\eqref}[1]{Eq.~\ref{eq:#1}}
\newcommand{\eqreftwo}[2]{Eqs.~\ref{eq:#1}--\ref{eq:#2}}
\newcommand{\Eqref}[1]{Equation~\ref{eq:#1}}
\newcommand{\figref}[1]{Fig.~\ref{fig:#1}}
\newcommand{\Figref}[1]{Figure~\ref{fig:#1}}

\newcommand{\secref}[1]{Sec.~\ref{sec:#1}}
\newcommand{\appref}[1]{Appendix~\ref{sec:#1}}
\newcommand{\appreftwo}[2]{Appendices~\ref{sec:#1}~and~\ref{sec:#2}}
\newcommand{\Secref}[1]{Section~\ref{sec:#1}}

\makeatother

\usepackage{babel}
\begin{document}

\title{Fluctuating surface-current formulation of radiative heat
  transfer: theory and applications}

\author{Alejandro W. Rodriguez}
\affiliation{School of Engineering and Applied Sciences, Harvard University, Cambridge, MA 02138}
\affiliation{Department of Mathematics, Massachusetts Institute of Technology, Cambridge, MA 02139}
\author{M. T. H. Reid}
\affiliation{Department of Mathematics, Massachusetts Institute of Technology, Cambridge, MA 02139}
\author{Steven G. Johnson}
\affiliation{Department of Mathematics, Massachusetts Institute of Technology, Cambridge, MA 02139}

\begin{abstract}
  We describe a novel fluctuating-surface current formulation of
  radiative heat transfer between bodies of arbitrary shape that
  exploits efficient and sophisticated techniques from the
  surface-integral-equation formulation of classical electromagnetic
  scattering. Unlike previous approaches to non-equilibrium
  fluctuations that involve scattering matrices---relating
  ``incoming'' and ``outgoing'' waves from each body---our approach is
  formulated in terms of ``unknown'' surface currents, laying at the
  surfaces of the bodies, that need not satisfy any wave equation. We
  show that our formulation can be applied as a spectral method to
  obtain fast-converging semi-analytical formulas in high-symmetry
  geometries using specialized spectral bases that conform to the
  surfaces of the bodies (e.g. Fourier series for planar bodies or
  spherical harmonics for spherical bodies), and can also be employed
  as a numerical method by exploiting the generality of surface
  meshes/grids to obtain results in more complicated geometries
  (e.g. interleaved bodies as well as bodies with sharp corners). In
  particular, our formalism allows direct application of the
  boundary-element method, a robust and powerful numerical
  implementation of the surface-integral formulation of classical
  electromagnetism, which we use to obtain results in new geometries,
  such as the heat transfer between finite slabs, cylinders, and
  cones.
\end{abstract}

\maketitle

\section{Introduction}

Quantum and thermal fluctuations of charges in otherwise neutral
bodies lead to stochastic electromagnetic (EM) fields everywhere in
space. In systems at equilibrium, these fluctuations give rise to
Casimir forces (generalizations of van~der~Waals interactions between
macroscopic bodies) that have recently become the subject of intense
theoretical and experimental
work.\cite{Genet08,Bordag09:book,Rodriguez11:review} In
non-equilibrium situations involving bodies at different temperatures,
these fields also mediate energy exchange from the hotter to the
colder bodies, a process known as \emph{radiative heat
  transfer}. Although the basic theoretical formalism for studying
heat transfer was laid out decades
ago~\cite{Rytov89,PolderVanHove71,Loomis94,Pendry99} only recently
have experiments reached the precision required to measure them at the
microscale,\cite{Kittel05,Narayanaswamy08,Hu08,Narayanaswamy09,Rousseau09,Sheng09,Wilde09,Ottens11}
sparking renewed interest in the study of these interactions in
complex geometries that deviate from the simple parallel-plate
structures of the
past.\cite{Mulet02,Joulain05,Chen05,Carey06,Fu06,Volokitin07,Zhang07,BasuZhang09}
In this manuscript, we present a novel formulation of radiative heat
transfer for arbitrary geometries based on the well-known
surface-integral-equation (SIE) formulation of classical
electromagnetism,\cite{Harrington89,Hackbush89,Rao99,bonnet99} which
extends our recently developed fluctuating surface-current (FSC)
approach to equilibrium Casimir forces~\cite{Reid12:FSC} to the
non-equilibrium problem of energy transfer between bodies of inequal
temperatures. Unlike scattering formulations based on basis expansions
of the field unknowns best suited to
special~\cite{Narayanaswamy08:spheres,bimonte09,Messina11,Kruger11,OteyFan11,Golyk12,Kruger12}
or non-interleaved periodic
\cite{bimonte09,Biehs11:apl,Guerout12,Marachevsky12} geometries, or
formulations based on expensive, brute-force time-domain
simulations\cite{RodriguezIl11} and Green's functions
calculations,~\cite{Volokitin01,Narayanaswamy13} this approach allows
direct application of the boundary element method (BEM): a mature and
sophisticated SIE formulation of the scattering problem in which the
EM fields are determined by the solution of an algebraic equation
involving a smaller set of surface unknowns (fictitious surface
currents in the surfaces of the
objects~\cite{Harrington89,Rao99,bonnet99}).

A terse derivation of our FSC formulation for heat transfer was
previously published in \citeasnoun{RodriguezReid12:FSC}. The primary
goals of this paper are to provide a more detailed presentation of
this derivation and to generalize our previous formula for the heat
transfer between two bodies to other situations of interest, including
geometries consisting of multiple and/or nested bodies. We also
demonstrate that the FSC framework can be applied as a spectral method
to obtain semi-analytical formulas in special geometries with high
symmetry, as well as for purely numerical evaluation using BEM, which
we exploit to obtain new results in a number of complicated geometries
that prove challenging for semi-analytical calculations. Although our
formulation here employs similar guiding principles as our previous
work on equilibrium Casimir phenomena~\cite{Reid11,Reid12:FSC}---both
are based on the SIE framework of classical EM scattering---the
heat-transfer case is by no means a straightforward extension of force
calculations, because generalizing the equilibrium framework to
non-equilibrium situations requires very different theoretical
techniques. For example, the fact that in \citeasnoun{Reid12:FSC} we
considered only \emph{equilibrium} fluctuations made it possible for
us to directly exploit the fluctuation-dissipation theorem for EM
fields,\cite{Eckhardt81} which relates the field--field correlation
function at two points to a \emph{single} Green's function between
those two points.  In contrast, although a fluctuation-dissipation
theorem exists in the non-equilibrium problem, the field--field
correlation functions are in this case determined by a product of two
Green's functions integrated over the volumes of the
bodies.\cite{Eckhardt81,Volokitin07} A key step in our derivation
below is a correspondence between this volume integral (involving
products of fields) and an equivalent surface integral involving the
fictitious surface currents and fields of the SIE framework, that was
not required in the equilibrium case.

The heat radiation and heat transfer of bodies with sizes and/or
separations comparable to the thermal wavelength can deviate strongly
from the predictions of the Stefan-Boltzmann
law.\cite{Rytov89,Eckhardt84} For instance, in the far field (object
separations $d$ much greater than the thermal wavelength $\lambda_T =
\hbar c / k_B T$), radiative heat transfer is dominated by the
exchange of propagating waves and is thus nearly insensitive to
changes in separations (oscillations from interference effects
typically being small~\cite{PolderVanHove71,Tschikin12}).  In the
less-studied near field regime ($d\lesssim \lambda_T$), not only are
interference effects important, but otherwise-negligible evanescent
waves also contribute flux through
tunneling.\cite{Zhang07,BasuZhang09} Such near-field effects have been
most commonly studied in planar geometries, where the monotonically
increasing contribution of evanescent waves with decreasing $d$
results in orders-of-magnitude enhancement of the net radiative heat
transfer rate (exceeding the far-field black-body limit at sub-micron
separations~\cite{BasuZhang09}). This enhancement was predicted
theoretically~\cite{PolderVanHove71,Zhang07,BasuZhang09} and observed
experimentally~\cite{Cravalho67,Hargreaves69,Domoto70} decades ago in
various planar structures, and has recently become the subject of
increased attention due to its potential application in
nanotechnology, with ramifications for thermal
photovoltaics~\cite{Pan00,Laroche06} and thermal
rectification,\cite{Chang06,Otey10,Basu11:apl,Iizuka12}
nanolithography,\cite{Lee08:heat} thermally assisted magnetic
recording,\cite{Challener09} and high-resolution surface
imaging.\cite{Kittel08,Wilde09,Biehs10:jap} Thus far, there have been
numerous works focused on the effects of material choice in planar
bodies,\cite{Fu06,Wang09} including studies of graphene
sheets,\cite{Ilic12} hyperbolic~\cite{Biehs12} and anisotropic
materials,\cite{Biehs11:apl} and even materials exhibiting phase
transitions,\cite{Zwol11} to name a few. Along the same lines, many
authors have explored transfer mediated by surface polaritons in thin
films\cite{Biehs07,Francoeur08,Fu09:heat,Basu11} and 1d-periodic
planar bodies.\cite{Ben-Abdallah10} Despite decades of research,
little is known about the near-field heat transfer characteristics of
bodies whose shapes differ significantly from these planar,
unpatterned structures. Theoretical calculations were only recently
extended to handle more complicated geometries, including
spheres,\cite{Narayanaswamy08:spheres,OteyFan11}
cylinders,\cite{Kruger11} and cones\cite{McCauleyReid12} suspended
above slabs, dipoles interacting with other
dipoles\cite{Chapuis08prb,Perez-Madrid09,Volokitin01,Domingues05,Perez-Madrid08,Perez-Madrid09,Manjavacas12}
or with surfaces,\cite{Huth10,Ben-Abdallah11,Tschikin12,Bellomo13} and
also patterned/periodic
surfaces.\cite{Biehs08,bimonte09,Ruting10,Messina11,RodriguezIl11,RodriguezReid12:FSC,RodriguezReid12:nonmon}

General-purpose methods for modelling heat transfer between bodies of
arbitrary shapes can be distinguished in at least two ways, in the
abstract \emph{formulation} of the heat-transfer problem and in the
\emph{basis} used to ``discretize'' the formulation into a finite
number of unknowns for solution on a computer (or by
hand).\cite{Johnson11:review} Theoretical work on heat transfer has
mainly centered on ``scattering-matrix'' formulations which express
the heat transfer in terms of the matrices relating incoming and
outgoing wave solutions from each
body.\cite{Biehs08,bimonte09,Messina11,Kruger11,Guerout12,Marachevsky12}
These formulations tend to be closely associated with ``spectral''
discretization techniques, in which a Fourier-like basis (Fourier
series, spectral harmonics, etc.)  is used to expand the unknowns,
because the incoming/outgoing waves must be expressed in terms of
known solutions of Maxwell's equations, which are typically a spectral
basis of planewaves, spherical waves, and so on.  Such a spectral
basis has the advantage that it can be extremely efficient
(exponentially convergent) if the basis is specially designed for the
geometry at hand (e.g. spherical waves for spherical
bodies\cite{Narayanaswamy08:spheres}).  Scattering-matrix methods can
also be used for arbitrary geometries, e.g. by expanding arbitrary
periodic structures in Fourier
series\cite{Biehs08,bimonte09,Guerout12} or by coupling to a generic
grid/mesh discretization to solve the scattering
problems,\cite{RodriguezIl11,RodriguezReid12:FSC,RodriguezReid12:nonmon}
but exponential convergence no longer generally obtains.  Furthermore,
Fourier or spherical-harmonic bases of incoming/outgoing waves
correspond to uniform angular/spatial resolution and require a
separating plane/sphere between bodies, which can be a disadvantage
for interleaved bodies or bodies with corners or other features
favoring nonuniform resolution.  In contrast to the geometric
specificity encoded in a particular scattering basis, one extremely
generic approach is a brute-force discretization of space and time,
allowing one to solve for heat-transfer by a Langevin
approach~\cite{RodriguezIl11} that handles all geometries equally,
including geometries with continuously varying material properties.
The FSC approach lies midway between these two extremes.  Like the
scattering-matrix approach, the FSC approach exploits the fact that
one knows the EM solutions (Green's functions) analytically in
homogeneous regions, so for piecewise-homogeneous geometries the only
remaining task is to match boundary conditions at interfaces.  Unlike
the scattering-matrix approach, however, the FSC approach is
formulated in terms of unknown surface currents rather than
incoming/outgoing waves---the surface currents are arbitrary vector
fields and need not satisfy any wave equation, which leads to great
flexibility in the choice of basis.  As described in this paper, the
FSC formulation can use either a spectral basis or a generic
grid/mesh and, as demonstrated in~\citeasnoun{RodriguezReid12:FSC} and
\citeasnoun{RodriguezReid12:nonmon}, works equally well for
interleaved bodies (lacking a separating plane or even a well-defined
notion of ``incoming/outgoing'' wave solutions).  Moreover, the FSC
formulation reduces the heat-transfer problem to a simple trace
formula in terms of well-studied matrices that arise in SIE
formulations of classical EM, which allows mature BEM solvers to be
exploited with minimal additional computational effort.

The radiative heat transfer between two bodies~1 and~2 at local
temperatures $T^\o{1}$ and $T^\o{2}$ can be written
as:\cite{Zhang07,BasuZhang09}
\begin{equation}
  H = \int_0^\infty d\omega \,
  \left[\Theta(\omega,T^\o{1})-\Theta(\omega,T^\o{2})\right]\Phi(\omega),
  \label{eq:H}
\end{equation}
where $\Theta(\omega,T) = \hbar\omega/[\exp(\hbar\omega/k_{B}T)-1]$ is
the Planck energy per oscillator at temperature~$T$, and $\Phi$ is an
ensemble-averaged \emph{flux spectrum} into body~2 due to random
currents in body~1 (defined more precisely below via the
fluctuation-dissipation
theorem~\cite{Lifshitz80,Rytov89,Eckhardt84}). (Physically, there are
currents in both bodies, but EM reciprocity~\cite{Jackson98} means
that one obtains the same $\Phi$ for flux into body~1 from sources in
body~2; this also ensures that $H$ obeys the second law of
thermodynamics.) The only question is how to compute $\Phi$, which
naively involves a cumbersome number of scattering calculations.

The main result of this manuscript is the compact trace-formula for
$\Phi$ derived in \secref{FSC-deriv}, which involves standard matrices
that arise in BEM calculations and forgoes any need for evaluation of
fields or sources in the volumes of the bodies, separation of incoming
and outgoing waves, integration of Poynting fluxes, or many scattering
calculations. As explained below in \secref{flux} and \secref{emm}, by
a slight modification of the two-body formula one can also
straightforwardly compute the spatially resolved pattern of Poynting
flux on the surfaces of the bodies, as well as the emissivity of an
isolated body. \Secref{heat-reciprocity} illustrates how important
physical properties such as reciprocity and positivity of heat
transfer manifest in the algebraic structure of the formulas. In
\secref{general}, we generalize the two-body formula to also describe
situations involving multiple and/or nested bodies.  The remaining
sections of the paper are devoted to validating the FSC formalism by
checking it against known results in special geometries consisting of
spheres and semi-infinite plates, as well as applying it to obtain new
results in more complicated geometries consisting of finite slabs,
cylinders, and cones. Specifically, \secref{spectral} considers
application of the FSC formulation in high-symmetry geometries where
the use of special-bases expansions involving Fourier and
spherical-wave eigenfunctions (provided in \appref{appendix-eigs})
leads to fast-converging semi-analytical formulas of heat radiation
and heat transfer for spheres and semi-infinite plates.  In
\secref{RWG} and \secref{apps}, we exploit a sophisticated numerical
implementation of the FSC formulation based on BEM to check the
predictions of the semi-analytical formulas in the case of spheres and
to obtain new results in more complex geometries. Finally, the
appendices at the end of the paper provide additional discussions that
supplement and aid our derivations in \secref{FSC-deriv} and
\secref{heat-transfer}. Specifically, \appref{appendix-equiv} provides
a concise derivation of the principle of equivalence and its
application to SIEs, and
\appreftwo{appendix-G-reciprocity}{appendix-BEM-reciprocity} provide
succinct proofs of reciprocity and positivity of Green's functions and
SIE matrices, respectively.

\section{FSC formulation}
\label{sec:FSC-deriv}

In this section, we review the SIE method of EM scattering and apply
it to derive an FSC formulation of radiative heat transfer between two
bodies. The result of this derivation is a compact trace expression
for $\Phi$ involving SIE matrices. We further elaborate on these
results in \secref{heat-transfer}, where we extend the formulation to
handle other situations of interest, including the emissivity of
isolated bodies, distribution of Poynting flux on the surfaces of the
bodies, and heat transfer between multiple and/or nested bodies.

\subsection{Notation}
\label{sec:notation}

Let $\f=\mat{\vec{E}\\\vec{H}}$ and $\s=\mat{\vec{J} \\ \vec{K}}$
denote 6-component volume electric and magnetic fields and currents,
respectively, and $\x$ denote 6-component \emph{surface} currents
(which technically have only 4 degrees of freedom since they are
constrained to flow tangentially to the surfaces).  In a homogeneous
medium, fields are related to currents via convolutions ($\star$) with
a $6\times 6$ homogeneous Green's tensor
$\G(\vec{x},\vec{y})=\G(\vec{x}-\vec{y},\vec{0})$, such that $\f = \G
\star (\s + \x)$, or more explicitly
\begin{equation}
  \f(\vec{x}) = \int d^3\vec{y} \, \G(\vec{x},\vec{y}) [\s(\vec{y})
  + \x(\vec{y})],
\end{equation}
where
\begin{equation*}
\G =
  \left(
  \begin{array}{cc}
    \vec{\G}^{EE} & \vec{\G}^{EH} \\
    \vec{\G}^{HE} & \vec{\G}^{HH}
  \end{array}
  \right)
  =  i k \left(
  \begin{array}{cc}
    Z\, \GD\xy & \CD\xy \\
    -\CD\xy & \frac{1}{Z} \GD \xy
  \end{array}
  \right)
\end{equation*}
is the Green's tensor composed of $3\times3$ electric and magnetic
Dyadic Green's functions (DGFs), determined by the ``photon'' DGFs
$\GD$ and $\CD$.  In the specific case of isotropic media (scalar
$\varepsilon$ and $\mu$), $\GD$ and $\CD$ satisfy
\begin{equation}
  \left[\nabla \times \nabla \times {} - k^2
    \right] \GD(k; \vec{x},\vec{x'}) =
  \delta(\vec{x}-\vec{x}')\mathbb{I},
\end{equation}
and $\CD = \frac{i}{k} \nabla \times \GD$, with wavenumber
$k=\omega\sqrt{\varepsilon\mu}$ and impedance $Z =
\sqrt{\mu/\varepsilon}$. Our derivation below applies to arbitrary
linear anisotropic permittivity $\varepsilon$ and permeability $\mu$,
so long as they are complex-symmetric matrices in order to satisfy
reciprocity~\cite{Landau:EM} (see \appref{appendix-G-reciprocity}).
The mathematical consequence of reciprocity, as described in the
Appendix, is that $\G$ is complex-symmetric up to sign flips.  In
particular, $\G(\vec{x},\vec{x}')^\T = \S \G(\vec{x},\vec{x}') \S$,
where the $6\times6$ matrix $\S = \S^{-1}$ flips the sign of the
magnetic components.  This reciprocity property is a key element of
our derivation below.

\subsection{Surface integral equations}
\label{sec:SIE}

\begin{figure}[t!]
\includegraphics[width=0.7\columnwidth]{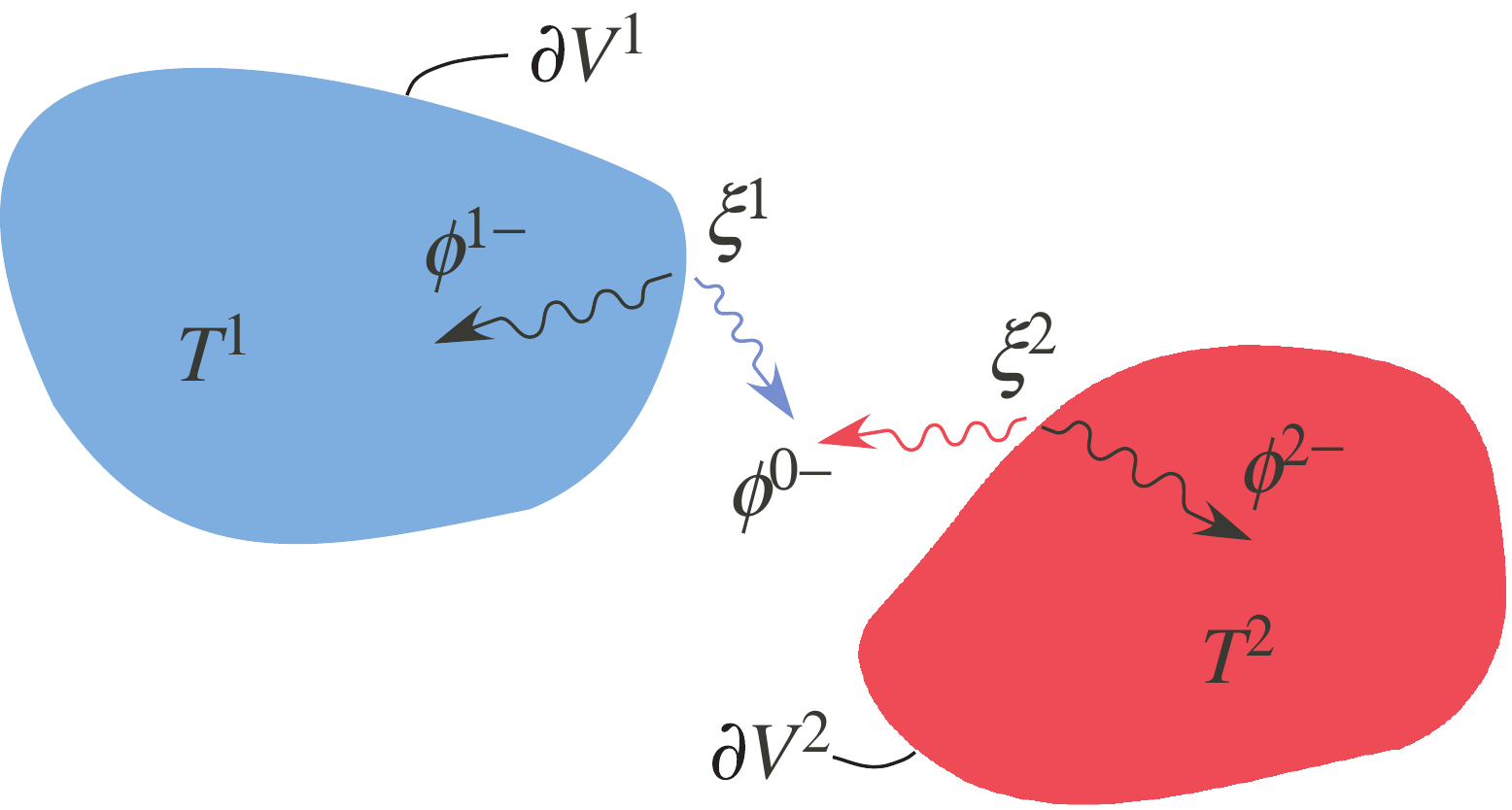}
\caption{Schematic depicting two disconnected bodies described by
  surfaces $\so^\o{1}$ and $\so^\o{2}$, and held at temperature
  $T^\o{1}$ and $T^\o{2}$, respectively. Surface currents $\xi^\o{1}$
  and $\xi^\o{2}$ laying on the surfaces of the bodies give rise to
  scattered fields $\f^\om{1}$ and $\f^\om{2}$, respectively, in the
  interior of the bodies, and scattered field $\f^\om{0}$ in the
  intervening medium 0.}
\label{fig:fig-twobodies}
\end{figure}

Consider the system depicted in \figref{fig-twobodies}, consisting of
two homogeneous bodies, 1 and 2 (volumes $V^\o{1}$ and $V^\o{2}$ and
temperatures $T^\o{1}$ and $T^\o{2}$), separated by a lossless medium
0 (volume $V^\o{0}$) by two interfaces $\so^\o{1}$ and $\so^\o{2}$,
respectively.  Consider also sources $\s^\o{r}$ located in the
interior of $V^\o{r}$, and denote the total fields in each region by
$\f^\o{r}$. The \emph{homogeneous}-medium Green's functions for the
infinite media in region $r$ are denoted by $\G^\o{r}$. Consider also
the decomposition of the total fields $\f^\o{r}$ in each region $r$
into ``incident'' fields $\f^\op{r}$ (due to sources within~$r$) and
``scattered'' fields $\f^\om{r}$ (from interactions with the other
regions, including both scattering off the interface and sources in
the other regions). That is, we can write: $\f^\o{r} = \f^\op{r} +
\f^\om{r},$ with $\f^\op{r} = \G^\o{r} \star \s^\o{r}$.

The core idea in the SIE formulation is the \emph{principle of
  equivalence},\cite{Love1901,Stratton39,Rengarajan00,Merewether80,Chen89,Harrington89}
whose derivation is briefly reprized in~\appref{appendix-equiv}, which
states that the scattered field $\f^\om{r}$ can be expressed as the
field of some \emph{fictitious} electric and magnetic surface currents
$\x^\o{r}$ located on the boundary of region~$r$, acting within an
infinite \emph{homogeneous} medium~$r$. In particular, one can write:
\begin{align}
  \f^\o{0} &= \f^\op{0} + \G^\o{0} \star (\x^\o{1} + \x^\o{2}) \\
  \f^\o{r} &= \f^\op{r} - \G^\o{r} \star \x^\o{r},
\end{align}
for $r=1,2$, with fictitious currents $\x^\o{r}$ completely determined
by the boundary condition of continuous tangential fields at the body
interfaces. Specifically, equating the tangential components of the
total fields at the surfaces of the bodies, we find the \emph{integral
  equations}:
\begin{equation}
  \left.(\G^\o{0}+\G^\o{r})\star\x^\o{r}+\G^\o{0}\star\x^{3-r}\right|_{\so^\o{r}}
  = \left.\f^\op{r}- \f^\op{0}\right|_{\so^\o{r}}
\label{eq:PMCHW}
\end{equation}
which can be solved to obtain $\x^\o{r}$ from the incident
fields. This is the ``PMCHW'' surface-integral formulation of EM
scattering.\cite{Harrington89,Umshankar86,Medgyesi94}

Let $\{\b_{n}^\o{r}\}$ be a \emph{basis} of 6-component tangential
vector fields on the surface of body~$r$, so that any surface current
$\x^\o{r}$ can be written in the form $\x^\o{r}(\vec{x}) =
\sum_{n}x_{n}^\o{r}\b_{n}^\o{r}(\vec{x})$ for $N$ coefficients
$\{x_{n}^\o{r}\}$. (In BEM, $\b_{n}$ is typically a
piecewise-polynomial ``element'' function defined within discretized
patches of each surface, most commonly the ``RWG'' basis
functions~\footnote{Typically, the surface-current vector fields are
  chosen to be either ``all-electric'' or ``all-magnetic'',
  i.e. $\beta_{nE}^r = \left(\begin{array}{c} \mathbf{b}_n \\
      0 \end{array}\right), \beta_{nM}^r = \left(\begin{array}{c} 0 \\
      \mathbf{b}_n \end{array}\right)$, where $\mathbf{b}_n$ is the
  usual 3-component RWG function corresponding to the $n$th internal
  edge in a surface mesh. However, in principle nothing precludes the
  use of mixed electric-magnetic surface-current basis
  functions}.\cite{Rao82} However, one could just as easily choose
$\b_{n}$ to be a spherical harmonic or some other ``spectral''
Fourier-like basis, as shown in \secref{spectral}. The key point is
that $\b_{n}$ is an arbitrary basis of surface vector fields; unlike
scattering-matrix formulations, it need \emph{not} consist of
``incoming'' or ``outgoing'' waves nor satisfy any wave equation.)
Taking the inner product of both sides of \eqref{PMCHW} with
$\b_m^\o{r}$ (a Galerkin discretization~\cite{boyd01:book}), one
obtains a matrix ``BEM'' equation of the form:
\begin{equation}
  W^{-1} x = s
\label{eq:BEM}
\end{equation}
where $x = \mat{x^\o{1} \\ x^\o{2}}$ represents the expansion of the
surface currents $\x^\o{r} = \sum_n x^\o{r}_n \b^\o{r}_n$, $s =
\mat{s^\o{1} \\ s^\o{2}}$ describes the effect of the incident fields
$s^\o{r}_m = \langle \b^\o{r}_m, \f^\op{r}-\f^\op{0} \rangle$, and
\begin{multline}
  \underbrace{\left(\begin{array}{cc}
      W^\o{11} & W^\o{12}\\
      W^\o{21} & W^\o{22}
    \end{array}\right)^{-1}}_{W^{-1}}=\\
    \underbrace{\left(\begin{array}{cc}
        G^\o{0,11} & G^\o{0,12} \\
        G^\o{0,21} & G^\o{0,22}
      \end{array}\right)}_{\hat{G}^\o{0}}+
    \underbrace{\left(\begin{array}{cc}
        G^\o{1}\\
        & 0
      \end{array}\right)}_{\hat{G}^\o{1}}+\underbrace{\left(\begin{array}{cc}
        0\\
        & G^\o{2}
      \end{array}\right)}_{\hat{G}^\o{2}}
\label{eq:W}
\end{multline}
describes interactions with matrix elements $G^\o{r,ij}_{mn} = \langle
\b^\o{i}_m, \G^\o{r} \star \b^\o{j}_n \rangle$ among basis
functions. $\hat{G}^\o{0}$ represents multi-body interactions between
basis functions on \emph{both} bodies, via waves propagating through
the intervening medium~0. $G^\o{r}$ represent self-interactions via
waves propagating within a body, given by:
\begin{align}
  G^\o{r}_{mn} \equiv G^\o{r,rr}_{mn} = \langle \b^\o{r}_m, \G^\o{r} \star \b^\o{r}_n \rangle .
  \label{eq:Gr}
\end{align}
Here, $\langle \cdot, \cdot \rangle$ denotes the standard inner
product $\langle \varphi, \psi \rangle = \int \varphi^* \psi$, with
the $*$ superscripts denoting the conjugate-transpose (adjoint)
operation.

A key property of the Green's function is reciprocity, as summarized
and derived in \appref{appendix-BEM-reciprocity}, and this property is
reflected in symmetries of the matrices $\hat{G}$ and $W$.  For
simplicity, let us begin by considering the case of real-valued basis
functions $\b_n$.  Let $S$ be the matrix such that $Sx$ flips the
signs of the magnetic components (assuming that we either have
separate basis functions for electric and magnetic components, as in
the RWG basis, or more generally that the basis functions come in
$\b_n$ and $\S \b_n$ pairs).  Note that $S^{-1} = S = S^\T = S^*$.  In
this case, as reviewed in \appref{appendix-BEM-reciprocity}, it
follows that $W^\T = SWS$ and $\hat{G}^\T = S\hat{G}S$.  Once we have
derived our heat-transfer formula for such real-valued basis
functions, it is straightforward to generalize to complex-valued bases
as described in \secref{complex}.

\subsection{Flux spectrum}
\label{sec:flux-spectrum}

Our goal is to compute the flux spectrum $\Phi$ into $V^\o{2}$ (the
absorbed power in body 2) due to dipole current sources $\s^\o{1}$ in
$V^\o{1}$ (integrated over all possible positions and
orientations). We begin by considering $\Phi_{\s^\o{1}}$, or the flux
into body 2 due to a \emph{single} dipole source $\s^\o{1}$ within
body 1, corresponding to $\f^\op{1} = \G^\o{1} \star \s^\o{1}$, with
$\f^\op{0} = \f^\op{2} = 0$. In the SIE \eqref{BEM}, this results in a
source term $s$ with $s^\o{1}_m = \langle \b^\o{1}_m, \G^\o{1} \star
\s^\o{1} \rangle$ and $s^\o{2} = 0$. As derived
in~\appref{appendix-equiv}, the Poynting flux can be computed using
the fact that $\x$ is actually equal to the surface-tangential
fields,\cite{Hackbush89} $\x = \begin{pmatrix} \vec{n} \times \vec{H}
  \\ -\vec{n}\times\vec{E} \end{pmatrix}$, where $\vec{n}$ is the
outward unit-normal vector.  It follows that the integrated flux
$-\frac{1}{2}\Re\oiint_2 (\overline{\vec{E}}\times\vec{H})\cdot\vec{n}
= \frac{1}{4} \Re \langle\x^\o{2},\f^\o{0}\rangle$.  (This can also be
derived as the power exerted on the surface currents by the total
field, with an additional $1/2$ factor from a subtlety of evaluating
the fields exactly on the surface.\cite{Chen89}) Hence:
\begin{equation*}
\Phi_{\s^\o{1}}=\frac{1}{4}\Re\langle\x^\o{2},\f^\o{0}\rangle=\frac{1}{4}\Re\langle\x^\o{2},\f^\o{2}\rangle=\frac{1}{4}\Re\langle\x^\o{2},-\G^\o{2}\star\x^\o{2}\rangle,
\end{equation*}
where we used the continuity of $\f^\o{0}$ and $\f^\o{2}$ and the fact
that $\f^\op{2}=0$. Substituting
$\x^\o{2}=\sum_{n}x_{n}^\o{2}\b_{n}^\o{2}$ and recalling the
definition of $G^\o{2}$ in \eqref{W}, we obtain
\begin{align*}
  \Phi_{\s^\o{1}} & = -\frac{1}{4}\Re\left(x^\os{2} G^\o{2} x^\o{2}\right) = -\frac{1}{4}\Re\left(x^{*}\hat{G}^\o{2}x\right) \\
  &=-\frac{1}{4} \left[x^* \left(\sym \hat{G}^\o{2} \right) x\right] 
  =-\frac{1}{4}s^{*}W^{*}\left(\sym\hat{G}^\o{2}\right)Ws \\
  & =-\frac{1}{4}\tr\left[ss^{*}W^{*}\left(\sym\hat{G}^\o{2}\right)W\right]
\end{align*}
where $\sym G = \frac{1}{2} (G + G^*)$ denotes the Hermitian
part of $G$.

Computing $\Phi_{\s^\o{1}}$ is therefore straightforward for a
single source $\s^\o{1}$. However, the total spectrum
\begin{equation}
  \Phi = \langle \Phi_{\s^\o{1}} \rangle =
  -\frac{1}{4}\tr\left[\langle ss^{*}\rangle
    W^{*}\left(\sym\hat{G}^\o{2}\right)W\right]
\label{eq:Phi_ens}
\end{equation}
involves an ensemble-average $\langle \cdots \rangle$ over all sources
$\s^\o{1}$ and polarizations in $V^\o{1}$. While this integration can
be performed explicitly, we instead seek to simplify matters so that
the final expression for $\Phi$ involves only surface integrals. The
key point is that $s s^*$ is an $N \times N$ matrix describing
interactions among the $N$ surface-current basis functions. The
ensemble average $ \langle s s^* \rangle$ is also an $N \times N$
matrix, which we would like to express in terms of a simple scattering
problem involving the SIE Green's function matrices, hence eliminating
any explicit computations over the interior volume $V^\o{1}$.

Defining the Hermitian matrix $\hat{C} = \langle s s^* \rangle$, it
follows that its only non-zero entries lie in the upper-left $N_1
\times N_1$ block $C^\o{1}= \langle s^1 s^{1*} \rangle$, and are given
by $C^\o{1}_{mn} = \langle s^1_m \overline{s^\o{1}_n} \rangle =
\left\langle \langle \b^\o{1}_m, \G^\o{1} \star \s^\o{1}\rangle
\langle \G^\o{1} \star \s^\o{1}, \b^\o{1}_n \rangle \right\rangle$, or
\begin{widetext}
\begin{align}
  C^\o{1}_{mn} &= \left\langle \oiint d^2\vec{x} \iiint d^3\vec{y}
  \b^\o{1}_m(\vec{x})^\T \G^\o{1}(\vec{x},\vec{y}) \s^\o{1}(\vec{y})
  \oiint d^2\vec{x}' \iiint d^3\vec{y}' \s^\o{1}(\vec{y}')^*
  \G^\o{1}(\vec{x}',\vec{y}')^* \b^\o{1}_n(\vec{x}') \right\rangle
  \nonumber\\ &= \oiint d^2\vec{x} \iiint d^3\vec{y} \b^\o{1}_m(\vec{x})^\T
  \G^\o{1}(\vec{x},\vec{y}) \oiint d^2\vec{x}' \iiint d^3\vec{y}'
  \Big\langle \s^\o{1}(\vec{y}) \s^\o{1}(\vec{y}')^* \Big\rangle
  \G^\o{1}(\vec{x}',\vec{y}')^* \b^\o{1}_n(\vec{x}') \nonumber\\ 
  &=
  \frac{4}{\pi} \oiint d^2\vec{x} \iiint d^3\vec{y} \oiint d^2\vec{x}'
  \b^\o{1}_m(\vec{x})^\T \G^\o{1}(\vec{x},\vec{y}) \left[\omega \Im
  \chi(\vec{y})\right] \G^\o{1}(\vec{x}',\vec{y})^* \b^\o{1}_n(\vec{x}'),
\label{eq:Cmn}
\end{align}
\end{widetext}
where in the third line we have performed an integration over all
dipole positions by employing the fluctuation-dissipation
theorem~\cite{Lifshitz80} for the current-current correlation
function,
\begin{equation}
  \langle \s^\o{1}(\vec{y}) \s^\o{1}(\vec{y}')^* \rangle =
  \frac{4}{\pi} \omega \Im \chi(\vec{y},\omega) \delta(\vec{y}-\vec{y}'),
\end{equation}
and where we omitted the dependence on the Planck energy distribution
$\Theta(\omega,T)$, which has been factored out into \eqref{H}, and
where $\Im \chi$ denotes the imaginary part of the $6\times 6$
material susceptibility tensor, so that $\Im\chi = \begin{pmatrix} \Im
  \varepsilon & 0 \\ 0 & \Im \mu \end{pmatrix}$, which is related to
material absorption.

\Eqref{Cmn} closely resembles an absorbed power in the volume of body
1, since absorbed power for a field $\f$ is $\frac{1}{2}\int \f^*
(\omega\Im\chi) \f$.\cite{Jackson98} To make this analogy precise,
some careful algebraic manipulation is required, and the
abovementioned reciprocity relations [$\G(\vec{x},\vec{x}')^\T = \S
  \G(\vec{x},\vec{x}') \S$, $W^\T = SWS$, etc] play a key role. In
particular, the fact that $C^\o{1}$ is Hermitian implies that the
matrix is completely determined by the values of $x^{\o{1}*} S
(C^\o{1})^\T S x^\o{1}$ for all $x^\o{1}$, where we have inserted the
sign-flip matrices $S$ and the transposition for later convenience.
Interpreting $x^\o{1}$ as the basis coefficients of a surface current
$\x^\o{1} = \sum_n x^\o{1}_n \b^\o{1}_n$ on $\so^\o{1}$, we find:
\begin{widetext}
\begin{align}
  x^{\o{1}*} S (C^\o{1})^\T S x^\o{1} &= \Big\langle \left|x^{\o{1}*}
  S \overline{s^\o{1}} \right|^2 \Big\rangle = \Big\langle
  \left|\x^\o{1}, S \overline{\G^\o{1} \star \s^\o{1}} \right|^2
  \Big\rangle \nonumber \\ &= \oiint d^2\vec{x} \iiint d^3\vec{y} \oiint
  d^2\vec{x}' \iiint d^3\vec{y}' \x^\o{1}(\vec{x})^{*} S
  \overline{\G^\o{1}(\vec{x},\vec{y})} \Big\langle
  \overline{\s^\o{1}(\vec{y})} \s^\o{1}(\vec{y}')^\T \Big\rangle
  \G^\o{1}(\vec{x}',\vec{y})^\T S \x^\o{1}(\vec{x}') \nonumber \\ &=
  \frac{4}{\pi} \oiint d^2\vec{x} \iiint d^3\vec{y} \oiint d^2\vec{x}'
  \x^\o{1}(\vec{x})^{*} \S \overline{\G^\o{1}(\vec{x},\vec{y})}
  \left[\omega \Im \chi(\vec{y})\right] \S \G^\o{1}(\vec{x}',\vec{y})
  \x^\o{1}(\vec{x}') \nonumber \\ &= \frac{4}{\pi} \oiint d^2\vec{x} \iiint
  d^3\vec{y} \oiint d^2\vec{x}' \x^\o{1}(\vec{x})^* \S
  \overline{\G^\o{1}(\vec{x},\vec{y})}\S \left[\omega \Im
    \chi(\vec{y})\right] \G^\o{1}(\vec{x}',\vec{y})
  \x^\o{1}(\vec{x}') \nonumber \\ &= \frac{4}{\pi} \oiint d^2\vec{x} \iiint
  d^3\vec{y} \oiint d^2\vec{x}' \left[\G^\o{1}(\vec{y},\vec{x})
    \x^\o{1}(\vec{x})\right]^{*} \left[\omega \Im
    \chi(\vec{y})\right] \left[\G^\o{1}(\vec{x}',\vec{y}) 
    \x^\o{1}(\vec{x}')\right] \nonumber \\ &= \frac{4}{\pi} \left\langle
  \G^\o{1} \star \x^\o{1}, (\omega \Im \chi) \G^\o{1} \star   \x^\o{1} \right\rangle 
\end{align}
\end{widetext}
where in the first and fourth lines we invoked reciprocity (from
above) and in the third line we assumed that $\S$ commutes with $\Im
\chi$, which is true for reciprocal media. (The only way that $\S$
would not commute with $\Im \chi$ would be if there were a chiral
susceptibility coupling electric and magnetic fields directly, also
called a bi-anisotropic susceptibility, which breaks
reciprocity~\cite{Lindell94,Serdyukov01}.)  Letting $\f^\o{1} =
\G^\o{1} \star \x^\o{1}$ be the field due to the surface current
$\x^\o{1}$, it follows that
\begin{equation}
  x^{\o{1}*} S (C^\o{1})^\T S x^\o{1} = \frac{4}{\pi} \langle \f^\o{1},
  (\omega \Im \chi) \f^\o{1} \rangle.
\label{eq:C1}
\end{equation}
But, as noted above, $\frac{1}{2} \langle \f^\o{1}, \omega (\Im \chi)
\f^\o{1}\rangle$ (where the inner product $\langle \cdot, \cdot
\rangle$ is now over the volume $V^\o{1}$) has a simple meaning: it is
the \emph{absorbed power} in $V^\o{1}$ from the currents $\x^\o{1}$,
or equivalently, the time-average power density dissipated in the
interior of body 1 by the field $\f^\o{1}$ produced by
$\x^\o{1}$.

Computing the interior dissipated power from an \emph{arbitrary}
surface current turns out to be somewhat complicated, since one needs
to take into account the possibility that the equivalent surface
currents arise from sources both outside and inside $V^\o{1}$.  If, on
the other hand, we could restrict ourselves to equivalent currents
$\xi^\o{1}$ which are outside of $V^\o{1}$, then we can use the result
from above that the incoming Poynting flux (the absorbed power) is
simply $- \frac{1}{4} \Re\langle\x^\o{1},\f^\o{1}\rangle = -\frac{1}{4}
x^\os{1} (\sym G^\o{1}) x^\o{1}$.  Substituting this into \eqref{C1},
we would be immediately led to the identity $x^{\o{1}*} S (C^\o{1})^\T
S x^\o{1} = -\frac{2}{\pi} \Re \,( x^\os{1} G^\o{1} x^\o{1})$, and
this gives an expression for $C^\o{1}$ in terms of $G^\o{1}$. It turns
out that indeed, we need not handle arbitrary $\xi^\o{1}$ since the
$\hat{C}$ matrix is never used by itself---it is only used in the
trace expression
\begin{align}
  \Phi &= -\frac{1}{4}\tr\left[\hat{C}W^{*}(\sym\hat{G}^\o{2})W\right] =
  -\frac{1}{4}\tr\left[\cdots\right]^\T \nonumber \\ &=
  -\frac{1}{4}\tr\left[S W S S (\sym \hat{G}^\o{2})S S
    W^{*} S \hat{C}^\T \right] \nonumber \\
  &= -\frac{1}{4}\tr[S W(\sym\hat{G}^\o{2})W^{*} S \hat{C}^T] \nonumber \\
  &= -\frac{1}{4}\tr[S \hat{C}^\T S W(\sym\hat{G}^\o{2})W^{*}],
\label{eq:PhiC}
\end{align}
using reciprocity. As shown in \secref{heat-reciprocity}, the standard
definiteness properties of the Green's functions (currents do
non-negative work) imply that $\sym\hat{G}^\o{r}$ is negative
semidefinite and hence admits a Cholesky
factorization~\cite{Trefethen97}
$\sym\hat{G}^\o{r}=-\hat{U}^\os{r}\hat{U}^\o{r}$. It follows that
\eqref{PhiC} can be written as $-\frac{1}{4}\tr[X^{*}S\hat{C}^\T SX]$,
where $X=W\hat{U}^\os{2}$ are the ``currents'' due to ``sources''
represented by the columns of $\hat{U}^\os{2}$, which are all of the
form $\mat{0 \\ s^\o{2}}$: currents from sources in $V^\o{2}$
alone. So, effectively, $S\hat{C}^\T S$ is only used to evaluate the
power dissipated in $V^\o{1}$ from sources in $V^\o{2}$, and by the
same Poynting-theorem reasoning from above, it follows that
$S(C^\o{1})^\T S=-\frac{2}{\pi}\sym G^\o{1}$, and hence
\begin{equation}
  \hat{C}=-\frac{2}{\pi}\sym S(\hat{G}^\o{1})^\T S
  =-\frac{2}{\pi}\sym\hat{G}^\o{1}
\label{eq:CsymG}
\end{equation}
by the symmetry of $\hat{G}^\o{1}$.  Substituting this result into
\eqref{Phi_ens} then gives the heat transfer formulation summarized in
the next section.

\subsection{Heat transfer formula}

The result of the above derivation is that the ensemble-averaged flux
from~$V^\o{1}$ to~$V^\o{2}$ can be expressed in the compact form:
\begin{empheq}[box=\fbox]{align}
  \Phi &= \frac{1}{2\pi} \tr \left[\left(\sym \hat{G}^\o{1}\right) W^*
    \left(\sym \hat{G}^\o{2}\right) W \right] \\
  &= \frac{1}{2\pi} \tr \left[\left(\sym G^\o{1}\right) W^\os{21}
    \left(\sym G^\o{2}\right) W^\o{21} \right],
\label{eq:Phi}
\end{empheq}
with $W^\o{21}$ relating incident fields at the surface of body~2 to
the equivalent currents at the surface of body~1. Our simplified
expression is computationally convenient because it only involves
standard matrices that arise in BEM calculations,\cite{Rao99} with no
explicit need for evaluation of fields or sources in the
volumes,\cite{Volokitin01,Narayanaswamy08:spheres,RodriguezIl11}
separation of incoming and outgoing
waves,\cite{Biehs08,bimonte09,Messina11,Kruger11,Guerout12,Marachevsky12}
integration of Poynting fluxes,\cite{RodriguezIl11} or any additional
scattering calculations.


\section{Generalizations}
\label{sec:heat-transfer}

In this section, we study the positivity and symmetries of the
two-body heat-transfer formula above and consider generalizations to
include other situations of interest. Following similar arguments as
those employed in the previous section, we derive formulas for the
emissivity of isolated bodies, the spatial distribution of Poynting
flux on the surfaces of bodies, and the heat transfer between multiple
and nested bodies. In \secref{complex}, we show that abandoning our
choice of real-$\b$ basis functions above in favor of complex-$\b$
functions does not change the final formula for $\Phi$, so long as the
$\b$s come in complex conjugate pairs.

\subsection{Positivity and reciprocity}
\label{sec:heat-reciprocity}

In addition to its computational elegance, \eqref{Phi} algebraically
captures crucial physical properties of the flux spectrum: $\Phi$ is
\emph{positive-definite} $\Phi \ge 0$ and symmetric with respect to
$1\leftrightarrow2$ exchange, as required by \emph{reciprocity}. Of
course, the positivity of $\Phi$ is immediately clear from the Rytov
starting point of fluctuating currents inside the bodies: the absorbed
power in one body from sources in the other body is simply $\sim \int
(\omega \Im \epsilon) |\vec{E}|^2 \geq 0$ (since $\omega\Im\epsilon
\geq 0$ for passive media~\cite{Landau:EM,Jackson98}).  Hence,
positivity must hold for any formulation that is mathematically
equivalent to the Rytov picture.  However, it is still useful and
nontrivial to understand how this positivity manifests itself
algebraically in a given formulation.  For example,
\citeasnoun{Kruger12} showed how positivity manifests itself in a
scattering-matrix framework.  In our FSC framework, positivity turns
out to correspond to the fact that $\Phi$ can be interpreted as kind
of matrix norm.

As derived above, the standard definiteness properties of the Green's
functions (currents do nonnegative work) imply that $\sym G^\o{r}$ is
negative semidefinite and hence admits a Cholesky factorization $\sym
G^\o{r}=-U^\os{r}U^\o{r}$, where $U^\o{r}$ is upper-triangular. It
follows that 
\begin{align}
  \Phi &= \frac{1}{2\pi} \tr \left[U^\o{1} W^{*} U^\os{2} U^\o{2} W
    U^\os{1}\right] \nonumber \\
  &= \frac{1}{2\pi}\tr[Z^{*}Z]=\frac{1}{2\pi}\Vert Z\Vert_{F}^\o{2},
\end{align}
where $Z=U^\o{2} W U^\os{1}$, is a weighted Frobenius norm of the SIE
matrix $W$, which from above we know is necessarily non-negative.

Furthermore, reciprocity (symmetry of $\Phi$ under $1\leftrightarrow2$
interchange) corresponds to simple symmetries of the matrices. As
derived in \appref{appendix-G-reciprocity},
$\G(\vec{y},\vec{x})^\T=\S\G(\vec{x},\vec{y})\S$,
$\hat{G}^\T=S\hat{G}S$ and $W^\T=S W S$, where $S=S^\T=S^{-1}=S^{*}$
is the matrix that flips the signs of the magnetic basis coefficients
and swaps the coefficients of $\b_{n}$ and $\overline{\b_{n}}$. It
follows that
\begin{align}
  \Phi & =\frac{1}{2\pi}\tr\left[S W S\left(\sym
    S\hat{G}^\o{2}S\right)S W^{*} S\left(\sym
    S\hat{G}^\o{1}S\right)\right]\nonumber \\ &
  =\frac{1}{2\pi}\tr\left[\left(\sym\hat{G}^\o{2}\right)W^{*}
    \left(\sym\hat{G}^\o{1}\right)W\right],
  \label{eq:Phi-symmetric}
\end{align}
where the $S$ factors cancel, leading to the $1\leftrightarrow2$
exchange.

\subsection{Complex-valued basis functions}
\label{sec:complex}

For convenience, we assumed above that the basis functions $\b_n$ were
purely real-valued.  However, it easy to generalize the final result
{\it a posteriori} to complex-valued basis functions.  The relevant
case to consider are basis functions that come in complex-conjugate
pairs $\b_n$ and $\b_{n'} =\overline{\b_n}$ (true for any practical
complex basis). Such a basis can always be transformed into an
equivalent real-valued basis $\tilde\b_n$ by the linear transformation
$\tilde\b_n = \frac{1}{\sqrt{2}}(\b_n + \b_{n'})$ and $\tilde\b_{n'} =
\frac{i}{\sqrt{2}}(\b_n - \b_{n'})$.  In an expansion $\xi = \sum_n
x_n \b_n = \sum_n \tilde{x}_n \tilde\b_n$, this is simply a rotation
$\tilde{x} = Qx$ where the matrix $Q$ is easily verified to be unitary
($Q^* = Q^{-1}$), since it is composed of unitary $2\times 2$ blocks
(operating on $n,n'$ complex-conjugate pairs).  Given such a unitary
change of basis, we can make a corresponding unitary change to the $G$
and $W$ matrices from above. $\tilde{G} = Q \hat{G} Q^*$ and $\tilde W
= Q W Q^*$, to obtain the matrices in the complex basis.  By
inspection of the $\Phi$ expression above, all of the $Q$ factors
cancel after the change of basis and one obtains the same expression
in the complex basis with the new $\tilde G$ and $\tilde W$ matrices.

\subsection{Emissivity of a single body}
\label{sec:emm}

The same formalism can be applied to compute the emissivity of a
single body. For a \emph{single} body~1 in medium~0, the emissivity of
the body is the flux $\Phi^\o{0}$ of random sources in~$V^\o{1}$
into~$V^\o{0}$.\cite{BasuZhang09} Following the derivation above, the
flux into~$V^\o{0}$ is
$-\frac{1}{4}\Re\langle\x^\o{1},\f^\o{0}\rangle=-\frac{1}{4}\langle\x^\o{1},\G^\o{0}\star\x^\o{1}\rangle$.
The rest of the derivation is essentially unchanged except that
$W=(G^\o{1}+G^\o{0})^{-1}$ since there is no second surface.  Hence,
we obtain
\begin{empheq}[box=\fbox]{equation}
\Phi^\o{0} = \frac{1}{2\pi}\tr\left[\left(\sym
  G^\o{1}\right)W^*\left(\sym G^\o{0}\right)W\right],
\label{eq:emm}
\end{empheq}
which again is invariant under $1\leftrightarrow 0$ interchange from
the reciprocity relations (Kirchhoff's law).

\subsection{Surface Poynting-flux pattern}
\label{sec:flux}

It is also interesting to consider the spatial distribution of
Poynting-flux pattern, which can be obtained easily because, as
explained above, $\frac{1}{4} \Re[\x^\o{2}(\vec{x})^*
  \f^\o{2}(\vec{x})]$ is exactly the inward Poynting flux at a point
$\vec{x}$ on surface~2. It follows that the mean contribution
$\Phi^\o{2}_n$ of a basis function $\b^\o{r}_n$ to $\Phi$ is
\begin{align*}
\Phi^\o{2}_n &= -\frac{1}{4} \left\langle \Re\left[ s^* W^* e^\o{2}_n
  e^\os{2}_n \hat{G}^\o{2} W s \right] \right\rangle \\ &=
-\frac{1}{4} \Re\left[ e^\os{2}_n\hat{G}^\o{2} W\langle s s^* \rangle
  W^* e^\o{2}_n \right] \\ &= \frac{1}{2\pi} \Re\left[
  e^\os{2}_n\hat{G}^\o{2} W \left(\sym \hat{G}^\o{1}\right) W^*
  e^\o{2}_n \right],
\end{align*}
where $e^\o{2}_n$ is the unit vector corresponding to the $\b^\o{2}_n$
component.  This further simplifies to $\Phi^\o{2}_n = F^\o{2}_{nn}$,
where
\begin{equation}
  F^\o{2} = \frac{1}{2\pi} \Re\left[ G^\o{2} W^\o{21}
    \left(\sym G^\o{1}\right) W^\os{21} \right].
\label{eq:flux}
\end{equation}
Note that $\Phi = \tr F^\o{2}$.  Similarly, by swapping
$1\leftrightarrow 2$ we obtain a matrix $F^\o{1}$ such that
$\Phi^\o{1}_n = F^\o{1}_{nn}$ is the contribution of $\b^\o{1}_n$ to
the flux on surface~$\so^\o{1}$.

\subsection{Multiple and nested bodies}
\label{sec:general}

In this section, we extend the FSC formalism above to situations
involving multiple and nested bodies. For simplicity, we only consider
an additional medium 3, since generalizations to include additional
bodies or levels of nesting readily follow.  Because the derivation is
almost identical to the two-body case, we only focus on those aspects
that differ.

\subsubsection{Multiple bodies}
\label{sec:many-bodies}

\begin{figure}[t!]
\includegraphics[width=0.7\columnwidth]{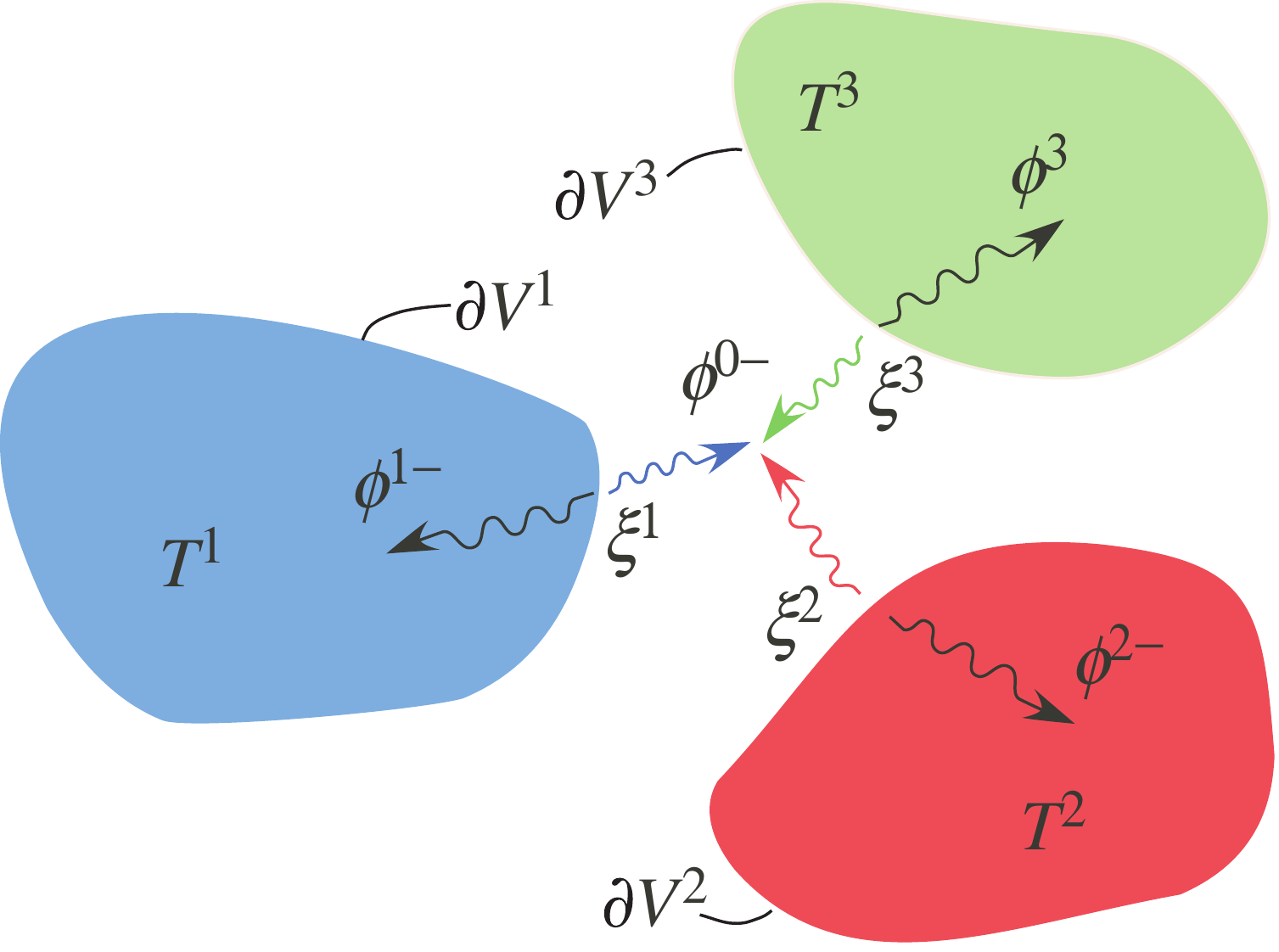}
\caption{Schematic depicting three disconnected bodies described by
  surfaces $\so^\o{1}$, $\so^\o{2}$, and $\so^\o{3}$, and held at
  temperature $T^\o{1}$, $T^\o{2}$, and $T^\o{3}$,
  respectively. Surface currents $\xi^\o{1}$, $\xi^\o{2}$, and
  $\xi^\o{3}$, laying on the surfaces of the bodies give rise to
  scattered fields $\f^\om{1}$, $\f^\om{2}$, and $\f^\om{3}$,
  respectively, in the interior of the bodies, and scattered field
  $\f^\om{0}$ in the intervening medium 0.}
\label{fig:fig-multbodies}
\end{figure}

Consider the system depicted in \figref{fig-multbodies}, consisting of
three disconnected bodies at different temperatures. Applying the
principle of equivalence, one finds:
\begin{align*}
  \f^\o{0} &= \f^\op{0} + \G^\o{0} \star (\x^\o{1} + \x^\o{2} + \x^\o{3}), \\
  \f^\o{r} &= \f^\op{r} - \G^\o{r} \star \x^\o{r},
\end{align*} 
for $r=1,2,3$, with fictitious currents $\x^\o{r}$ determined by the
boundary conditions of continuous tangential fields at the the body
interfaces. Equating the tangential components of the fields at the
surfaces of the bodies, one obtains the integral equations:
\begin{equation}
  (\G^\o{0}+\G^\o{r})\star\x^\o{r}+ \sum_{i\neq r} (\G^\o{0} \star
  \x^\o{j}) \,\big|_{\so^\o{r}} = \left.\f^\op{r}-
  \f^\op{0}\right|_{\so^\o{r}},
\end{equation}
along with the corresponding SIE matrix:
\begin{multline}
  \underbrace{\mat{W^\o{11} & W^\o{12} & W^\o{13} \\ 
      W^\o{21} & W^\o{22} & W^\o{23} \\ 
      W^\o{31} & W^\o{32} & W^\o{33}}^{-1}}_{W^{-1}} = 
  \underbrace{\mat{G^\o{0,11} & G^\o{0,12} & G^\o{0,13} \\
      G^\o{0,21} & G^\o{0,22} & G^\o{0,23} \\
      G^\o{0,31} & G^\o{0,32} & G^\o{0,33}}}_{\hat{G}^\o{0}} 
  +
  \\
  \underbrace{\mat{G^\o{1} & & \\
      & 0 & \\
      & & 0}}_{\hat{G}^\o{1}}
  +
  \underbrace{\mat{0 & & \\
      & G^\o{2} &  \\
      &  & 0}}_{\hat{G}^\o{2}}
  +
  \underbrace{\mat{0 & & \\
      & 0 & \\
      & & G^\o{3}}}_{\hat{G}^\o{3}}.
\end{multline}
The derivation of the flux spectrum for any given pair of bodies
mirrors \emph{exactly} the derivation in \secref{FSC-deriv}, with the
only difference being the modified SIE matrix $W$. The final
expression for the flux spectrum into $V^\o{j}$ due to random currents
in $V^\o{i\neq j}$ is given by:
\begin{equation}
  \Phi^\o{ij} = \frac{1}{2\pi} \tr \left[ (\sym G^\o{i}) W^\os{ji}
    (\sym G^\o{j}) W^\o{ji} \right],
\end{equation}
which again is invariant under $i \leftrightarrow j$ interchange.

\subsubsection{Nested bodies}
\label{sec:nested-bodies}

\begin{figure}[t!]
\includegraphics[width=0.7\columnwidth]{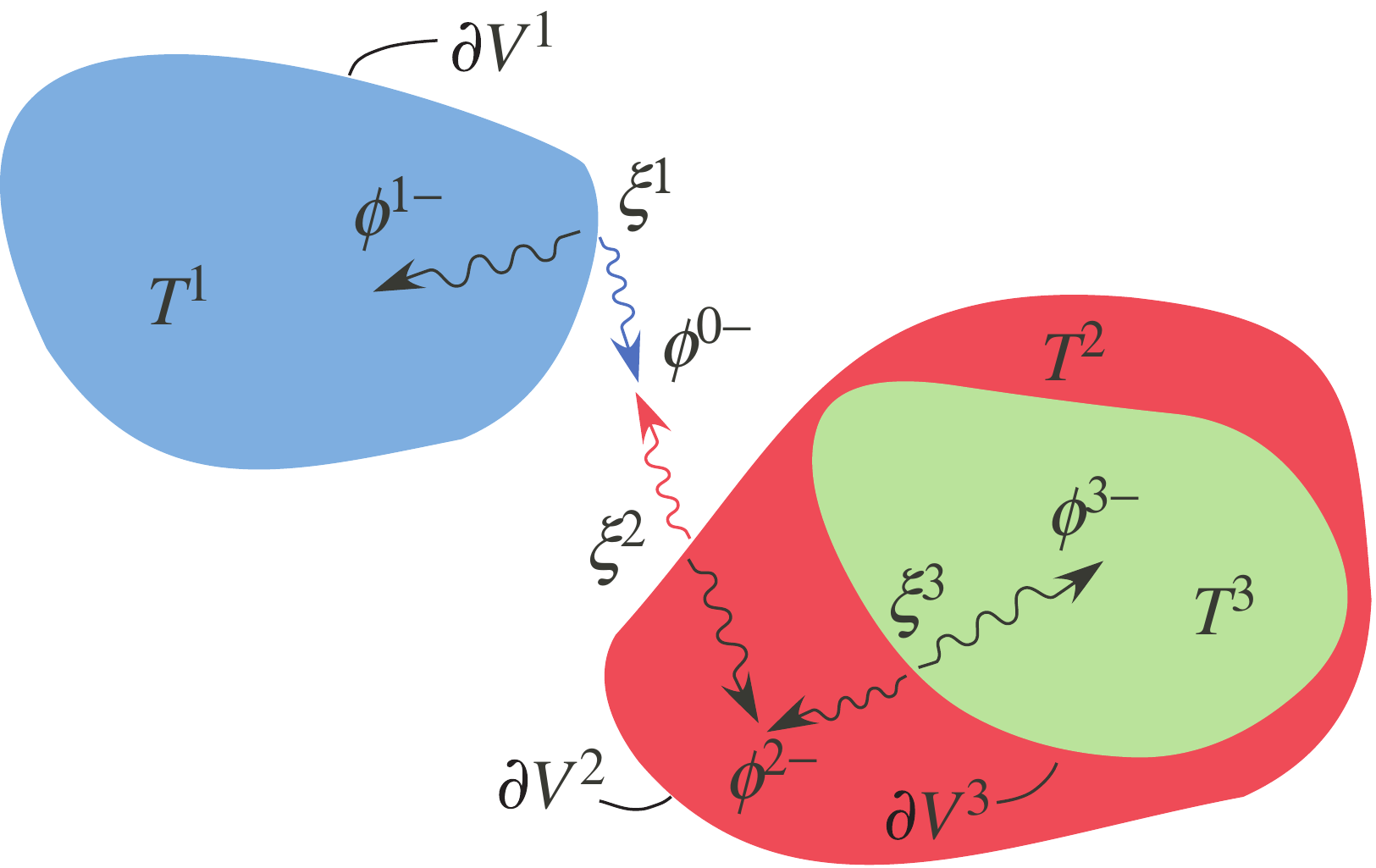}
\caption{Similar three-body geometry as that depicted in
  \figref{fig-multbodies}, but with body 3 now embedded in the
  interior of body 2. Here, the scattered field $\f^\om{2}$ includes
  contributions from both $\xi^\o{2}$ and $\xi^\o{3}$.}
\label{fig:fig-nestedbodies}
\end{figure}

Consider now the system depicted in \figref{fig-nestedbodies},
involving three bodies at different temperatures with one of the
bodies (medium 2) containing another (medium 3). Applying the
principle of equivalence again, one finds:
\begin{align*}
  \f^\o{0} &= \f^\op{0} + \G^\o{0} \star (\x^\o{1} + \x^\o{2}), \\
  \f^\o{2} &= \f^\op{2} - \G^\o{2} \star (\x^\o{2} - \x^\o{3}), \\
  \f^\o{r} &= \f^\op{r} - \G^\o{r} \star \x^\o{r},
\end{align*}
for $r=1,3$, with fictitious currents $\xi^\o{r}$ determined by the
boundary conditions of continuous tangential fields at the body
interfaces. Equating the tangential components of the fields at the
the surfaces of the bodies, one obtains the integral equations:
\begin{align*}
  \left. (\G^\o{0} + \G^\o{1}) \star \x^\o{1} + \G^\o{0} \star \x^\o{2} \right|_{\so^\o{1}} 
  &= \left. \f^\op{1}-\f^\op{0} \right|_{\so^\o{1}}, \\ 
  \left. (\G^\o{0} +  \G^\o{2}) \star \x^\o{2} + \G^\o{0} \star \x^\o{1} - \G^\o{2}
  \star \x^\o{3} \right|_{\so^\o{2}} 
  &= \left. \f^\op{2}-\f^\op{0} \right|_{\so^\o{2}}, \\ 
\left.  (\G^\o{2} + \G^\o{3}) \star \x^\o{3} - \G^\o{2} \star \x^\o{2} \right|_{\so^\o{3}} 
  &= \left. \f^\op{3}-\f^\op{2} \right|_{\so^\o{3}},
\end{align*}
where $\partial V^\o{2}$ denotes the interface between $V^\o{2}$ and
$V^\o{0}$, from which one obtains the corresponding SIE matrix:
\begin{multline}
  \underbrace{\mat{W^\o{11} & W^\o{12} & W^\o{13} \\ 
      W^\o{21} & W^\o{22} & W^\o{23} \\ 
      W^\o{31} & W^\o{32} & W^\o{33}}^{-1}}_{W^{-1}} = 
  \underbrace{\mat{G^\o{0,11} & G^\o{0,12} & \\
      G^\o{0,21} & G^\o{0,22} & \\
      & & 0}}_{\hat{G}^\o{0}} 
  +
  \\
  \hspace{0.15in}
  \underbrace{\mat{G^\o{1} & & \\
      & 0 & \\
      & & 0}}_{\hat{G}^\o{1}}
  +
  \underbrace{\mat{0 & & \\
     & G^\o{2} & -G^\o{2,23} \\
      & -G^\o{2,32} & G^\o{2,33}}}_{\hat{G}^\o{2}}
  +
  \underbrace{\mat{0 & & \\
      & 0 & \\
      & & G^\o{3}}}_{\hat{G}^\o{3}} .
\end{multline}
Although the derivation of the flux spectrum for any given pair of
bodies closely mirrors the derivation in \secref{flux-spectrum},
important deviations arise due to the difference in topology. In what
follows, we only focus on those steps that differ significantly. The
asymmetry of the geometry also means that we must consider $\Phi$ for
each pair separately.

First, we compute the flux spectrum $\Phi^\o{13}$ into $V^\o{3}$ (the
absorbed power in 3) due to dipole current sources in $V^\o{1}$. The
flux into body 3 due to a single dipole source $\s^\o{1}$ inside body
1 is given by:
\begin{align*}
  \Phi^\o{13}_{\s^\o{1}} &= \frac{1}{4} \Re \langle \x^\o{3},
  \f^\o{2}\rangle = \frac{1}{4} \Re \langle \x^\o{3},
  \f^\o{3}\rangle \nonumber \\ &= \frac{1}{4} \Re \langle
  \x^\o{3},-\G^\o{3} \star \f^\o{3}\rangle \nonumber \\ &=
  -\frac{1}{4} \Re \left(x^\os{3}G^\o{3} x^\o{3}\right)
\end{align*}
After ensemble averaging over $\s^\o{1}$ as before, we obtain:
\begin{align}
  \Phi^\o{13} &= \frac{1}{2\pi} \tr \left[(\sym G^\o{1}) W^\os{31}
    (\sym G^\o{3}) W^\o{31}\right].
\label{eq:Phi13}
\end{align}
Second, we compute the flux spectrum $\Phi^\o{12}$ into $V^\o{2}$ (the
absorbed power in body 2) due to dipole current sources in $V^\o{1}$.
Direct application of Poynting's theorem at $\so^\o{2}$ in this case
does \emph{not} yield $\Phi^\o{12}$ but rather the quantity we denote
as $\Phi^\o{1(2)}$: the flux into the entire region contained by
$\so^\o{2}$ from sources in $V^\o{1}$, which includes absorption in
both $V^\o{2}$ \emph{and} $V^\o{3}$. It follows that $\Phi^\o{12} =
\Phi^\o{1(2)} - \Phi^\o{13}$.  So, it only remains to compute
$\Phi^\o{1(2)}$, starting with the flux from a single $\s^\o{1}$
source, given by:
\begin{align*}
  \Phi^\o{1(2)}_{\s^\o{1}} &= \frac{1}{4} \Re \langle \x^\o{2},
  \f^\o{0}\rangle = \frac{1}{4} \Re \langle \x^\o{2}, \f^\o{2}\rangle 
\nonumber \\ &=
  \frac{1}{4} \Re\langle \x^\o{2}, -\G^\o{2} \star
  (\x^\o{2}-\x^\o{3})\rangle
 \nonumber \\ &=
  -\frac{1}{4} \Re \left[ x^\os{2} \left(G^\o{2} x^\o{2} -
      G^\o{2,23} x^\o{3}\right) \right],
\label{eq:Phi12}
\end{align*}
with the additional $x^\o{3}$ term stemming from absorbed power in
body~3.  We ensemble average as before, and obtain
\begin{align}
  \Phi^\o{12} &= \Phi^\o{1(2)} - \Phi^\o{13} \nonumber \\
&= \frac{1}{2\pi} \tr \Big[ (\sym G^\o{1}) W^\os{21}
    (\sym G^\o{2}) W^\o{21} \nonumber \\ &\hspace{0.2in} - (\sym G^\o{1})
    \sym\left(W^\os{21} G^\o{2,23} W^\o{31}\right)\Big]
- \Phi^\o{13} .
\end{align}
Finally, we compute the flux spectrum $\Phi^\o{32}$ into $V^\o{2}$
(the absorbed power in body 2) due to dipole current sources in
$V^\o{3}$. $\Phi^\o{23}$ can be computed by subtracting the flux
$\Phi^\o{3(2)}$ leaving body 2 through $\so^\o{2}$ from the flux
$\Phi^\o{3(3)}$ entering body 2 through $\so^\o{3}$. Specifically, for
a single dipole $\s^\o{3}$, we find:
\begin{align*}
  \Phi^\o{32}_{\s^\o{3}} &= \frac{1}{4} \Re \langle \x^\o{3},
  \f^\o{2}\rangle - \frac{1}{4} \Re \langle \x^\o{2}, \f^\o{2}\rangle
  \\ &= \frac{1}{4} \Re \langle \x^\o{3}, \f^\o{3}\rangle -
  \frac{1}{4} \Re \langle \x^\o{2}, \f^\o{2}\rangle \nonumber \\ &=
  \frac{1}{4} \Re\langle \x^\o{3}, -\G^\o{3} \star \x^\o{3}\rangle -
  \frac{1}{4} \Re \langle \x^\o{2}, -\G^\o{2} \star
  (\x^\o{2}-\x^\o{3})\rangle \nonumber \\ &= \underbrace{-\frac{1}{4}
    \Re \left(x^\os{3} G^3 x^\o{3}\right)}_{\Phi^\o{3(3)}_{\s^\o{3}}} +
  \underbrace{\frac{1}{4} \Re \left[ x^\os{2} \left(G^\o{2} x^\o{2} -
      G^\o{2,23} x^\o{3}\right) \right]}_{-\Phi^\o{3(2)}_{\s^\o{3}}}.
\label{eq:Phi23}
\end{align*}
The final result is the expression:
\begin{equation}
 \Phi^\o{32} = \Phi^\o{3(3)} - \Phi^\o{3(2)},
\end{equation}
with
\begin{align}
  \Phi^\o{3(2)} 
  &= \frac{1}{2\pi} \tr \left[ (\sym G^\o{3}) W^\os{23} (\sym G^\o{2}) W^\o{23} \right. \nonumber \\
    &\left. \hspace{0.6in} + \, (\sym G^\o{3}) \sym \left( W^\os{23} G^\o{2,23} W^\o{33}\right)\right] \\
  \Phi^\o{3(3)} 
  &= \frac{1}{2\pi} \tr \left[ (\sym G^\o{3}) W^\os{33}
    (\sym G^\o{3}) W^\o{33}\right].
\end{align}

For example, the heat transfer between $V^\o{1}$ and the combined $V^\o{(2)} = V^\o{2} \cup V^\o{3}$ is given by
\begin{equation}
   H^\o{1(2)} = \int \Theta_{T^\o{1}} \Phi^\o{1(2)} -
    \Theta_{T^\o{2}} \Phi^\o{12} - \Theta_{T^\o{3}} \Phi^\o{13},
\end{equation}
where the integral is taken over all positive frequencies $\omega$ and
$\Theta_T \equiv \Theta(\omega,T)$. In the special case $T^\o{2} =
T^\o{3}$, the expression reduces to the expected form:
\begin{equation}
   H^\o{1(2)} = \int \left(\Theta_{T^\o{1}}-\Theta_{T^\o{2}}\right)
    \Phi^\o{1(2)}
\end{equation}
As before, we obtain reciprocity relations $\Phi^\o{ij} = \Phi^\o{ji}$
between every pair of bodies, but these relations are no longer
apparent merely by inspection of $\Phi^\o{ij}$.  Because each body is
topologically distinct, $\Phi^\o{ji}$ is no longer obtained from
$\Phi^\o{ij}$ merely by interchanging $i$ and $j$, but instead must be
derived separately (using analogous steps).  Upon carrying out this
derivation, we verify that $\Phi^\o{ij} = \Phi^\o{ji}$ as
required. Furthermore, the positivity of $\Phi^\o{ij}$ appears harder
to derive algebraically from the final expression than in the
non-nested cases, and we do not do so in this work.  (Although it
follows from the second law of thermodynamics, the scattering-matrix
proof of positivity~\cite{Kruger12} should apply to nested bodies with
minimal modification).

\section{Validation}
\label{sec:validation}

We now apply our FSC formulation to obtain results obtained previously
using other scattering formulations in several high-symmetry
geometries. In \secref{basischoice}, we discuss the choice of basis,
contrasting BEMs that use a generic surface mesh with spectral methods
that use a Fourier-like basis, and point out that the latter are
actually closely related to scattering-matrix methods in the case of
high-symmetry geometries.  In \secref{spectral}, we derive
semi-analytical expressions of heat radiation and heat transfer for
spheres and plates, using surface spherical-harmonics and Fourier
bases to describe the SIE surface unknowns, and show that these agree
with previous formulae derived using other
formulations.\cite{Kattawar70,Narayanaswamy08,Golyk12,Kruger12} In
\secref{RWG}, we present a general-purpose numerical implementation of
the FSC formulation based on a standard triangular-mesh discretization
of the surfaces of the bodies known as the BEM ``RWG'' method; we
check it against previous heat-transfer methods by computing the heat
transfer between spheres.

\subsection{Choice of basis}
\label{sec:basischoice}

The standard approach for solving the SIEs above is to
\emph{discretize} them by introducing a finite set of basis functions
$\b_{n}$ defined on the surfaces of the bodies. As noted above, an
important property of SIE formulations is that $\b_{n}$ is an
arbitrary basis of surface vector fields: unlike scattering-matrix
formulations,\cite{bimonte09,Kruger11,Messina11} they need \emph{not}
satisfy any wave equation, nor encapsulate any global information
about the scattering geometry, nor consist of ``incoming'' or
``outgoing'' waves into or out of the bodies. This lack of restriction
on $\b_{n}$ is a powerful property of the SIE formalism.

There are two main categories of basis functions that one could
employ: \emph{spectral} bases or \emph{boundary-element} bases.  A
spectral basis consists of a Fourier-like complete basis of
non-localized functions, such as spherical harmonics or Chebyshev
polynomials,\cite{boyd01:book} which are truncated to obtain a finite
basis.  BEMs instead first discretize each surface into a mesh of
polygonal \emph{elements} (e.g. triangles), and describe functions
piecewise by low-degree polynomials in each
element.\cite{Harrington89,Hackbush89,bonnet99} Spectral bases have
the advantage that they can converge exponentially fast for smooth
functions,\cite{boyd01:book} or in this case for smooth interfaces,
but they are not as well suited to handle singularities such as
corners, and moreover represent surfaces with essentially uniform
spatial resolution.  A BEM basis, on the other hand, is more flexible
because it can use a nonuniform mesh to concentrate spatial resolution
where it is needed,\cite{Hackbush89,Rao99} and furthermore the
localized nature of the basis functions has numerical advantages in
assembling and applying the $W$ and $G$ (Green's function)
matrices.\cite{chew97,chew01} The most common BEM technique employs a
mesh of triangular elements (\emph{panels}) with vector-valued
polynomial basis functions called an RWG (Rao--Wilson--Glisson)
basis,\cite{Rao82} where each basis function is associated with each
\emph{edge} of the mesh and is nonzero over a pair of triangles
sharing that edge.  Many years of research have been devoted to the
efficient assembly of the $G$ matrices for the RWG basis (by
evaluating the singular panel integrals of
$\G$),\cite{Cai02,Taylor03,Tong06} and to fast methods for solving the
resulting linear equations.\cite{barrett94,chew01}

For a handful of highly symmetric geometries, however, spectral bases
have an additional advantage: a special basis can be chosen such that
most of the matrix elements can be computed analytically (and many of
the $G$ matrices are diagonal as a consequence of orthogonality).
This has a close connection to scattering-methods, because whenever
there is a known incoming/outgoing wave basis (e.g. spherical waves),
one can construct an equivalent set of surface-current basis functions
(e.g. spherical harmonics) by the principle of equivalence.  (In fact,
the principle of equivalence can be used to derive an \emph{exact}
equivalence between our $\Phi$ expressions and the analogous
expressions from the scattering-matrix formulation, which we do not
show here.)  In the example of interactions between two spherical
bodies, if we employ a (vector) spherical-harmonic basis on each body,
then the $G^\o{r}$ self-interaction matrices are diagonal and the
$G^\o{0,rr'}$ interaction matrix is given by ``translation matrices''
that relate spherical-wave bases at different
origins.\cite{Rahi09:PRD}  In this way, by choosing a
geometry-specific basis, the FSC formulation can retain all of the
efficiency of the scattering-matrix methods, while preserving the
flexibility to employ a different basis as needed.

\subsection{Spectral basis}
\label{sec:spectral}

In this section, we explicitly apply our FSC formulation with a
spectral basis in three high-symmetry geometries for which the matrix
elements can be evaluated semi-analytically: radiation of an isolated
plate and an isolated sphere, and heat transfer between two parallel
plates.  In each case, we reproduce known solutions that were derived
previously using scattering-matrix
formulations.\cite{Kattawar70,Narayanaswamy08,Golyk12,Kruger12}  The
main purpose of this section is to illustrate how the FSC formulation
with a spectral basis allows semi-analytical calculations similar to
scattering-matrix formulations (albeit only in the handful of
high-symmetry geometries where exact wave solutions can be constructed
in each body).  To begin with, we review the well-known spectral
representation of the homogeneous DGF $\G$ in bases specialized to
particular coordinate systems.

\subsubsection{Basis of Helmholtz solutions}

We wish to work with solutions of Maxwell's equations known
analytically within each body and which are orthogonal when evaluated
\emph{on the interfaces}. These solutions, evaluated at the interface
of each body, will then provide a basis of surface-tangential vector
fields in which the $G$ matrices can be evaluated analytically or
semi-analytically.  In particular, we wish to work with solutions
$\vec{M}$ and $\vec{N}$ of the vector Helmholtz equation (equivalent
to Maxwell's equations in a homogeneous isotropic
medium),\cite{Tau93}
\begin{equation}
  \Big[\nabla^2 + k^2 \Big] \mat{\vec{M} \\ \vec{N}} = 0,
\label{eq:Helmholtz}
\end{equation}
with $\vec{M} = -i/k \nabla \times \vec{N}$ and $\vec{N} =
i/k\nabla\times \vec{M}$ denoting purely electric and purely magnetic
vector fields. (Note that $\vec{M}$ and $\vec{N}$ come in two flavors,
depending on whether on solves \eqref{Helmholtz} for outgoing or
incoming boundary conditions.)  Furthermore, we seek solutions of
\eqref{Helmholtz} in a coordinate system that allows separation of
variables into ``normal'' and ``tangential'' components to some
surface $\so$ (which is possible for a small number of coordinate
systems). We let $\eta_\perp$ represent the separable coordinate
identified as the normal coordinate, and let
$\boldsymbol{\eta}_\parallel$ represent the remaining tangential
coordinates. The choice of coordinate system ultimately corresponds to
a choice of basis, or independent solutions labeled by an index $n$
that correspond to different scattering channels. Specifically, one is
led to vector fields:\cite{Tau93}
\begin{align}
\vec{M}^\pm_n(\eta_\perp,
\boldsymbol{\eta}_\parallel) &= \kappa_{n,E}^\pm(\eta_\perp)
\vec{X}_n(\boldsymbol{\eta}_\parallel) \\
\vec{N}^\pm_n(\eta_\perp, \boldsymbol{\eta}_\parallel) &=
\kappa_{n,M}^\pm(\eta_\perp)
\vec{Z}_n(\boldsymbol{\eta}_\parallel),
\end{align}
with $\kappa^\pm_n$ and $\{\vec{X}_n, \vec{Z}_n\}$ denoting the normal
and tangential components of the fields, and with $\pm$ denoting
incoming ($+$) and outgoing ($-$) solutions.  For example, solutions
in spherical coordinates yield the well-known vector spherical-wave
solutions $\vec{M}^\pm_{\ell,m}(r,\theta,\phi) = R^\pm_{\ell}(r)
\vec{Y}_{\ell,m}(\theta,\phi)$, described by spherical Hankel
functions $\kappa^\pm_{\ell,m,E} = R^\pm_\ell$ and vector spherical
harmonics $\vec{X}_{\ell,m} = \vec{Y}_{\ell,m}$ in terms of radial and
angular coordinates $\eta_\perp=r$ and $\boldsymbol{\eta}_\parallel =
\{\theta,\phi\}$, respectively, and labeled by angular-momentum
``quantum'' numbers $n = \{\ell,m\}$.

Because $\vec{M}_n$ and $\vec{N}_n$ form an orthonormal basis (due to
the self-adjointness of the Helmholtz operator), the homogeneous
photon DGFs $\GD$ and $\CD$ of \secref{notation} can be expressed in
such a basis as:\cite{Pathak83,Tau93,Rahi09:PRD}
\begin{widetext}
\begin{eqnarray}
  \label{eq:GD}
  \GD(k;\vec{x},\vec{x}') &=& \frac{\eta_\perp(\vechat{x})
    \eta_\perp(\vechat{x}')}{2 i k} \, \delta(\vec{x}-\vec{x}') +
  \sum_{n}
  \begin{cases}
    \chi_{n,E} \vec{M}^{+}_{n}(\vec{x}) \otimes
    \vec{M}^{-}_{n}(\vec{x}') + \chi_{n,M}
    \vec{N}^{+}_{n}(\vec{x})\otimes\vec{N}^{-}_{n}(\vec{x}')
    & \eta_\perp(\vec{x}) >
      \eta_\perp(\vec{x}')\\ \chi_{n,E}
      \vec{M}^{-}_{n}(\vec{x}) \otimes
      \vec{M}^{+}_{n}(\vec{x}') + \chi_{n,M}
      \vec{N}^{-}_{n}(\vec{x})
      \otimes\vec{N}^{+}_{n}(\vec{x}') 
      & \eta_\perp(\vec{x}) < \eta_\perp(\vec{x}'),
  \end{cases}
\end{eqnarray}
\end{widetext}
and $\CD = \frac{i}{k} \nabla \times \GD$, respectively, where the
coefficients $\chi_n$ are determined by taking the Wronskian of the
outgoing ($-$) and incoming ($+$) solutions.

The SIE matrices appearing in \eqref{Phi} involve inner products of
\eqref{GD} with basis functions $\b_n$ defined at the \emph{surfaces}
of the bodies. (Note that because the Green's functions are evaluated
on the surface, inclusion of the delta-function term is
crucial~\footnote{Mathematically, the $\delta$ function terms arising
  in the spectral expansion of the homogeneous DGF arise because the
  vector-valued basis functions $\vec{M}$ and $\vec{N}$ only span the
  transverse part of the space, and have no components along the
  longitudinal directions.\cite{Tau93}}.)  Just as the vector fields
$\vec{M}_n$ and $\vec{N}_n$ form a convenient basis in which to expand
waves propagating inside and outside $\so$, so too do the
\emph{tangential} components $\vec{X}_n$ and $\vec{Z}_n$ form a
suitable basis in which to express surface--current basis functions
$\b_n$ defined on $\so$. In this case, as in the case of RWG basis
functions,\cite{Rao99} $\b_n$ can be chosen to be purely electric
($E$) or purely magnetic ($M$), so that
\begin{equation}
  \b_{n,E} = \mat{\vec{X}_{n} \\ 0}, \,\,\,
  \b_{n,M} = \mat{0 \\ \vec{Z}_{n}}.
\end{equation}
Moreover, the orthogonality relations of the tangential vector fields,
$\langle \vec{X}_m, \vec{X}_n \rangle = \langle \vec{Z}_m, \vec{Z}_n
\rangle = \delta_{mn}$ and $\langle \vec{X}_m, \vec{Z}_n \rangle = 0$,
mean that only basis functions with the same $n$ and same polarization
have non-zero overlap. These surface currents form a complete basis
and satisfy convenient orthogonality relations with the corresponding
vector fields:
\begin{align}
  \label{eq:XM}
  \langle \vec{X}_m, \vec{M}_n^\pm \rangle &= \left(\left. \kappa_{n,E}^\pm
  \right|_{\so}\right) \delta_{mn} \\ \langle \vec{Z}_m, \vec{N}_n^\pm
  \rangle &= \left(\left. \kappa_{n,M}^\pm \right|_{\so}\right)
  \delta_{mn} \\ \langle \vec{X}_m, \vec{N}_n^\pm \rangle &= \langle
  \vec{Z}_m, \vec{M}_n^\pm \rangle = 0,
  \label{eq:XN}
\end{align}
with inner products $\langle \cdot, \cdot \rangle$ corresponding to
surface integrals over the tangential coordinates evaluated at the
surface $\so$, i.e. $\langle \varphi, \psi \rangle = \oiint_{\so}
d^2\boldsymbol{\eta}_\parallel \,
\mathcal{J}(\eta_\perp,\boldsymbol{\eta}_\parallel) \varphi^* \psi$,
where $\mathcal{J}$ denotes the Jacobian factor for the coordinate
system.

The combination of these orthogonality relations and the Green's
function expression of \eqref{GD}, implies that the $G$ matrices
arising in the SIE formulation for interface $\so$ will be diagonal
and known analytically in this basis.  Therefore, choosing this basis
simplifies the calculation of $\Phi$, as illustrated in the next
sections.

\subsubsection{Heat transfer formulation}

Expression of the homogeneous Green's function in the interior of each
high-symmetry body $r$ in the basis specialized for that body yields a
block-diagonal self-interaction matrix $G^\o{r}$ with matrix elements
$G^\o{r}_{mn,QQ'} = \langle \b^\o{r}_{m,Q}, \G^\o{r} \star\,
\b^\o{r}_{n,Q'} \rangle \sim \delta_{mn}$, where $Q$ denotes
polarization. In contrast, the lack of any orthogonality relations
between wave solutions constructed for different, unrelated bodies
means that the interaction matrices $\hat{G}^\o{0,rr'}$ are dense,
i.e. the matrix elements $G^\o{0,rr'}_{mn,QQ'} = \langle
\b^\o{r}_{m,Q}, \G^\o{0} \star \b^\o{r'}_{n,Q'} \rangle$ generally do
not vanish.  The outgoing fields into the intervening medium
$\G^\o{r} \star\, \b^\o{r}_{m,Q}$ due to currents in body $r$ are
still known analytically from \eqref{GD}, described in terms of the
wave solutions specialized to body $r$ (albeit evaluated in the
exterior medium $0$), but in order to take inner products with
$\b^\o{r'}_{n,Q}$ for a body $r'$ we need to ``translate'' the
solutions centered on $r$ to the different basis of waves centered on
$r' \neq r$.  Such change of bases are often performed via
``translation'' and ``conversion'' matrices that are well-known and
tabulated for most shapes of interest,\cite{Rahi09:PRD} and
immediately yield the SIE interaction matrices $G^\o{0,rr'}$.

For the remainder of the section, we restrict ourselves to situations
involving either a single body or two identical bodies described by a
common set of basis functions, in which case the individual SIE
matrices are block-diagonal in $n$ and polarization. In particular,
the $G$ matrices for a given $n$ are given by:
\begin{align*}
  \hat{G}^0 &= \left(
  \begin{array}{cccc}
    G^{0,11}_\perp & G^{0,12}_\perp & & \\
    G^{0,21}_\perp & G^{0,22}_\perp & & \\
    & & G^{0,11}_\parallel & G^{0,12}_\parallel \\
    & & G^{0,21}_\parallel & G^{0,22}_\parallel \\
  \end{array}
  \right), \\
  \hat{G}^1  &= 
  \left(
  \begin{array}{cccc}
    G^{1}_\perp & & & \\
    & 0 & & \\
    & & G^{1}_\parallel & \\
    & & & 0 \\
  \end{array}
  \right), \,\,\,
  \hat{G}^2  = 
  \left(
  \begin{array}{cccc}
    0 & & & \\
    & G^{2}_\perp & & \\
    & & 0 & \\
    & & & G^{2}_\parallel \\
  \end{array}
  \right)
\end{align*}
where $G_\perp$ and $G_\parallel = G_\perp (E \to H)$ are $2\times 2$
block matrices
\begin{eqnarray}
\label{eq:Gperp}
G_{\perp,nn} &=& 
  \left(
    \begin{array}{cc}
      \langle\vec{X}_{n},\vec{\G}^{EE}\star\vec{X}_{n}\rangle &
      \langle\vec{X}_{n},\vec{\G}^{EH}\star\vec{Z}_{n}\rangle \\
      \langle\vec{Z}_{n},\vec{\G}^{HE}\star\vec{X}_{n}\rangle &
      \langle\vec{Z}_{n},\vec{\G}^{HH}\star\vec{Z}_{n}\rangle
    \end{array}
    \right).
\label{eq:Gpar}
\end{eqnarray}
Here, the subscripts $\perp$ and $\parallel$ refer to the two
decoupled polarization states, corresponding to purely electric $E$
and purely magnetic $M$ surface currents, respectively. The
separability of the two polarizations means that the flux spectrum
$\Phi$ can be written in the form $\Phi=\sum_{\pol} \Phi_{\pol}$, with
$\Phi_\pol$ denoting the contribution of the $\pol$ polarization. From
the definitions of the $\G$ functions, it follows that the two are
related to one another by $\Phi_\parallel = \Phi_\perp (Z \to 1/Z)$.

In the subsequent sections, we derive semi-analytical expressions for
$\Phi$ in special geometries involving isolated and interacting plates
and spheres. The symmetry of these geometries make it convenient to
represent the SIE matrices using Fourier and spherical-wave surface
basis functions, described in \appref{appendix-eigs}. Our final
expressions agree with previous formulae derived using the
scattering-matrix
approach.\cite{Kattawar70,Narayanaswamy08,Golyk12,Kruger12}

\subsubsection{Isolated plates}

We first consider the radiation of an isolated plate. Using the
appropriate Fourier basis supplied in \appref{appendix-Fourier} and the
corresponding Green's function expansion of \eqref{GD}, the $G_\perp$
matrices for the plate are given by:
\begin{equation}
  G^{0,11}_\perp =
  \frac{1}{2} \left(
    \begin{array}{cc}
      \frac{Z_0}{\gamma_0} & 1 \\
      1 & -\frac{\gamma_0}{Z_0} 
    \end{array}
  \right),\,\,
  G^{1}_\perp =
  \frac{1}{2} \left(
    \begin{array}{cc}
      \frac{Z_1}{\gamma_1} & -1 \\
      -1 & -\frac{\gamma_1}{Z_1} 
    \end{array}
  \right)
\label{eq:plateG}
\end{equation}
where $\gamma_r = \sqrt{1-(|\vec{k}_\perp|/k^\o{r})^2}$ is the
wavenumber in the $z$ direction normalized by $k^\o{r}$.  It follows
that the flux spectra for the two polarizations are given by:
\begin{align}
  \Phi_{\perp} &= \frac{1}{4\pi} \mathrm{Tr} \left[
    \frac{\Re\left(\frac{\gamma_0}{Z_0}\right)
      \Re\left(\frac{\gamma_1}{Z_1}\right)}{\left|
      \frac{\gamma_0}{Z_0} + \frac{\gamma_1}{Z_1} \right|^2}\right],
  \,\, \Phi_{\parallel} = \Phi_{\perp} \left( Z \to \frac{1}{Z}
  \right),
\end{align}
with $\tr \Phi = \int \frac{d^2\vec{k}_\perp}{(2\pi)^2} \,
\Phi(\vec{k}_\perp)$ corresponding to integration over the parallel
wavevector. Assuming a non-dissipative external medium ($\Im
\varepsilon_0 = \Im \mu_0 = 0$), and performing straightforward
algebraic manipulations, one obtains the well-known formula for the
emissivity of the plate:\cite{Joulain05}
\begin{equation}
  \Phi(\omega) = \frac{1}{8\pi} \int_0^{\omega}
  \frac{d^2\vec{k}_\perp}{(2\pi)^2} \, \sum_{\pol=\{\perp,||\}}
  \epsilon_{\pol}(\vec{k}_\perp, \omega),
\label{eq:platephi}
\end{equation}
where $\epsilon_{\pol} = \frac{1}{2} (1 - |r_{\pol}|^2)$ denotes the
directional emissivity of the plate for the $\pol$ polarization,
expressed in terms of the Fresnel reflection
coefficients:\cite{Jackson98}
\begin{align}
  r_{\perp} &= \frac{\frac{\gamma_0}{Z_0} -
    \frac{\gamma_1}{Z_1}}{\frac{\gamma_0}{Z_0} + \frac{\gamma_1}{Z_1}}, \,\,\,
  r_{\parallel} = r_{\perp}\left(Z \to \frac{1}{Z}\right)
\label{eq:fresnel}
\end{align}

\subsubsection{Isolated spheres}

We now consider the radiation of an isolated sphere. Using the
appropriate vector spherical wave basis supplied in
\appref{appendix-mult} and the corresponding Green's function
expansion, the $G_\perp$ matrices for the sphere are given by:
\begin{align}
  G^{0,11}_\perp &= (z_0 R)^2 \left(
    \begin{array}{cc}
      Z_0 j_\ell(z_0) h_\ell(z_0) & i j_\ell(z_0) \breve{h}_\ell(z_0) \\ 
      -i j_\ell(z_0) \breve{h}_\ell(z_0) &
      \frac{1}{Z_0}\breve{j}_\ell(z_0)\breve{h}_\ell(z_0)
    \end{array}
  \right) \\
  G^{1}_\perp &= (z_1 R)^2 \left(
    \begin{array}{cc}
      Z_1 j_\ell(z_1) h_\ell(z_1) & i \breve{j}_\ell(z_1) h_\ell(z_1) \\ 
      -i \breve{j}_\ell(z_1) h_\ell(z_1) &
      \frac{1}{Z_1}\breve{j}_\ell(z_1)\breve{h}_\ell(z_1)
    \end{array}
  \right),
\label{eq:sphereG}
\end{align}
where $\breve{f}(z) \equiv (1/z + d/dz) f$, $j_\ell$ and $h_\ell$ are
Bessel functions of the first and second kind, respectively, and $z_r
= k^r R$. Employing a number of well-known properties of spherical
Bessel functions, such as the Wronskian identity $j_\ell'(z) h_\ell(z)
- h_\ell'(z) j_\ell(z) = i/z^2$, one arrives at the following flux
spectra for the two polarizations:
\begin{align}
  \Phi_\perp &= \frac{1}{8\pi} \mathrm{Tr} \left[\frac{1}{|z_0
      h_\ell(z_0)|^2} \frac{\Im \left[\frac{Z_0}{Z_1}
        \frac{\breve{j}_\ell(z_1)}{j_\ell(z_1)}\right]}{\left|\frac{Z_0}{Z_1}
      \frac{\breve{j}_\ell(z_1)}{j_\ell(z_1)} -
      \frac{\breve{h}_\ell(z_0)}{h_\ell(z_0)}\right|^2}\right] \\
  \Phi_\parallel &= \Phi_\perp \left(Z\to \frac{1}{Z}\right)
\end{align}
with $\tr \Phi = \sum_{\ell,m} \Phi_{\ell m}$ corresponding to a sum
over the angular-momentum quantum numbers.  Assuming vacuum as the
external medium ($\varepsilon_0 = \mu_0 = 1$) and a non-magnetic
sphere ($\mu_1=1$), one obtains the well-known formula for the
emissivity of a sphere in vacuum:\cite{Kattawar70}
\begin{multline}
  \Phi(\omega) = \frac{1}{8\pi} \sum_{\ell>1} \frac{(2\ell+1)}{|z_0
    h_\ell(z_0)|^2} \\ \left[ \frac{\Im \left[n_1
        \frac{\breve{j}_\ell(z_1)}{j_\ell(z_1)}\right]}{\left|n_1
      \frac{\breve{j}_\ell(z_1)}{j_\ell(z_1)} -
      \frac{\breve{h}_\ell(z_0)}{h_\ell(z_0)}\right|^2} + \frac{\Im
      \left[n_1^*
        \frac{\breve{j}_\ell(z_1)}{j_\ell(z_1)}\right]}{\left|n_1
      \frac{\breve{h}_\ell(z_0)}{h_\ell(z_0)} -
      \frac{\breve{j}_\ell(z_1)}{j_\ell(z_1)} \right|^2}\right],
\label{eq:spherephi}
\end{multline}
where $n_1 = \sqrt{\varepsilon_1}$ is the index of refraction of the
sphere.

\subsubsection{Two plates}

Finally, we consider the heat transfer between two parallel,
semi-infinite plates separated by distance $d$. Just as in the case of
isolated plates, it is convenient to express the $G_\perp$ matrices in
the Fourier basis supplied in \appref{appendix-Fourier}. Here, in
addition to the self-interaction matrices
\begin{align}
  G^\o{0,rr}_\perp = \frac{1}{2} \left(
    \begin{array}{cc}
      \frac{Z_0}{\gamma_0} & 1 \\
      1 & -\frac{\gamma_0}{Z_0} 
    \end{array}
  \right),\,\,
  G^{r}_\perp =
  \frac{1}{2} \left(
    \begin{array}{cc}
      \frac{Z_r}{\gamma_r} & -1 \\
      -1 & -\frac{\gamma_r}{Z_r} 
    \end{array}
  \right),
\end{align}
for $r=1,2$, one obtains the interaction or ``translation'' matrices
\begin{equation}
  G^\o{12}_\perp = G^\o{21}_\perp = 
  \frac{1}{2}
  \left(
  \begin{array}{cc}
      \frac{Z_0}{\gamma_0} & 1 \\
      1 & \frac{\gamma_0}{Z_0}     
  \end{array}
  \right) e^{i k_0 \gamma_0 d},
\end{equation}
where the exponential factors above couple or ``translate'' waves
arising in different origins. Straightforward matrix algebra yields
the following flux spectra for the two polarizations:
\begin{align}
  \Phi_\perp &= \frac{1}{2\pi} \mathrm{Tr}
  \left[\frac{\left|\frac{\gamma_0}{Z_0}e^{2ik_0\gamma_0
        d}\right|^2}{|\rho_\perp|^2} \frac{
        \Re\left(\frac{\gamma_1}{Z_1}\right)
        \Re\left(\frac{\gamma_2}{Z_2}\right)}{\left|\frac{\gamma_0}{Z_0}+\frac{\gamma_1}{Z_1}\right|^2\left|\frac{\gamma_0}{Z_0}+\frac{\gamma_2}{Z_2}\right|^2}
      \right] \\ \Phi_\parallel &= \Phi_\perp\left(Z\to
    \frac{1}{Z}\right),
\end{align}
where $\rho_{\pol} = |1 - r^1_{\pol} r^2_{\pol} e^{2i k_0 \gamma_0
  d}|^2$ and $r^q_{\pol}$ is the Fresnel reflection coefficient of
plate $q$ for the $\pol$ polarization given in
\eqref{fresnel}. Assuming a non-dissipative external medium ($\Im
\varepsilon_0 = \Im \mu_0 = 0$), and performing straightforward
algebraic manipulations, one obtains the well-known
formula:\cite{Joulain05}
\begin{equation}
   \Phi(\omega) = \Phi_{\mathrm{prop}}(\omega) 
   + \Phi_{\mathrm{evan}}(\omega),
\end{equation}
with
\begin{align}
\label{eq:Phi-plates}
  \Phi_{\mathrm{prop}}(\omega) &= \frac{1}{4\pi}
  \sum_{\pol} \int_0^{\omega}
  \frac{d^2\vec{k}_\perp}{(2\pi)^2} \, \frac{\epsilon^1_{\pol} 
    \epsilon^2_{\pol}}{\rho_{\pol}}  \\
  \Phi_{\mathrm{evan}}(\omega) &= \frac{1}{4\pi}
  \sum_{\pol} \int_\omega^{\infty}
  \frac{d^2\vec{k}_\perp}{(2\pi)^2}\, (\Im r^1_{\pol})(\Im r^2_{\pol})
  \frac{e^{-2\Im(k_0\gamma_0) d}}{\rho_{\pol}}  
\end{align}
where $\epsilon^q_{\pol}$ denotes the emissivity of plate $q$ for the
$\pol$ polarization, and where $\Phi$ has been conveniently decomposed
into far-field (propagating) and near-field (evanescently decaying)
contributions.

\subsection{BEM discretization via RWG basis}
\label{sec:RWG}

In contrast to spectral methods, BEMs discretize the surfaces of the
bodies into polygonal \emph{elements} or ``panels'', and describe
piecewise functions in each element by low-degree
polynomials.\cite{Hackbush89,bonnet99} The most common BEM technique
employs a so-called RWG basis of vector-valued polynomial functions
defined on a mesh of triangular panels.\cite{Rao82} Such a basis is
applicable to arbitrary geometries and yields results that converge
with increasing resolution (smaller triangles), where variants with
different convergence rates depend upon the degree of the polynomials
used in the triangles (which can be curved).  The simplest
discretizations involve degree-1 polynomials and flat triangles, where
the error decreases at least linearly with $1/$diameter of the
triangles, but can converge faster with adaptive mesh
refinements~\cite{bonnet99}. In contrast to spectral methods, the
$G_{nm}$ integrals here must be performed numerically and the
resulting $G$ matrices are dense, but thankfully fast techniques to
perform these integrals are well established and need only be
implemented once for a given RWG basis, independent of the
geometry.\cite{Rao82,Harrington89,Hackbush89} One such implementation
is the free-software solver \emph{\sc scuff-em}~\cite{Reid:scuffem},
which we exploit in this section to compare results from BEM RWG to
known results for spheres; the same code is employed in \secref{apps}
to obtain results in new and more complex geometries.

\begin{figure}[t!]
\includegraphics[width=1.0\columnwidth]{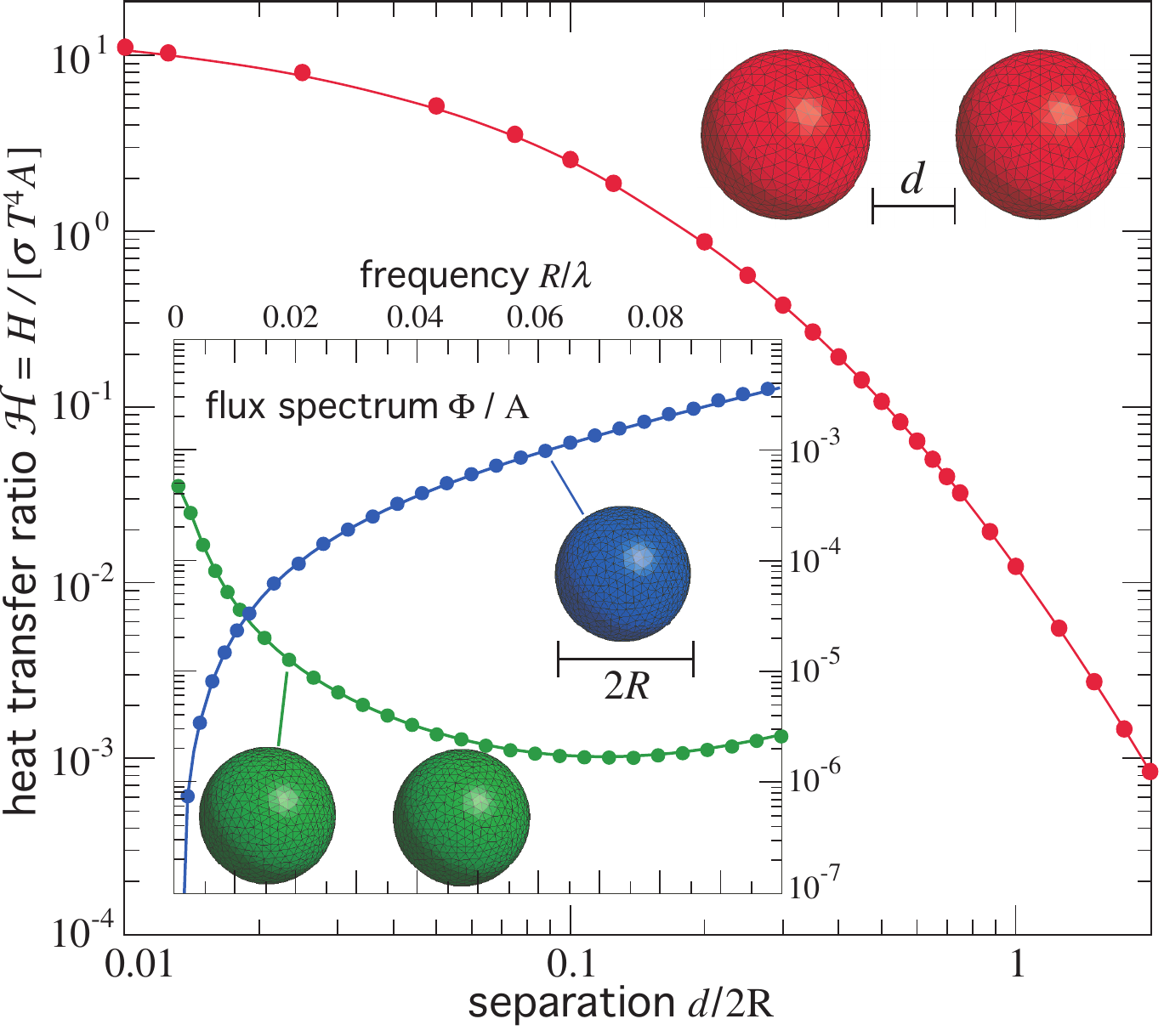}
\caption{Ratio $\h = H/\sigma T^4 A$ of the heat-transfer rate $H$
  between gold spheres of radii $R=0.2\mu$m and the Stefan--Boltzmann
  law $\sigma T^4 A$, where $A=4\pi R^2$ is the surface area of the
  spheres, with one sphere held at $T=300$~K and the other held at
  zero temperature, as a function of their surface--surface
  separation. (Inset:) Flux spectra $\Phi(\omega)$ per unit area $A$
  (units of $\mu\mathrm{m}^{2}$), of the two spheres at $d=R$ (green
  circles) and of an isolated sphere (blue circles).}
\label{fig:fig-spheres}
\end{figure}

The heat-transfer rate $H$ between two spheres was recently obtained
numerically by \citeasnoun{Narayanaswamy08:spheres}. In contrast to
scattering-matrix methods or the FSC formalism above, the method
of~\citeasnoun{Narayanaswamy08:spheres} involves straightforward
integration of the inhomogeneous Green's function of the geometry over
the volumes of the two spheres, expressed in terms of a specialized
spherical-wave basis expansion with coefficients determined by
enforcing continuity of the fields across the various interfaces. The
result of the integration is an exponentially convergent
semi-analytical formula of the kind derived in
\secref{spectral}. \Figref{fig-spheres} compares the results of the
BEM RWG method (red circles) against those obtained by evaluating the
semi-analytical formula of~\citeasnoun{Narayanaswamy08:spheres},
truncated at a sufficiently large but finite order (solid lines). In
particular, the heat transfer ratio $\h = H / \sigma T^4 A$ is plotted
as a function of surface--surface separation $d$ for gold spheres of
radius $R=1\mu$m, where one sphere is held at $T=300$~K while the
other is held at zero temperature, and where $\sigma T^4 A$ is the
Stefan--Boltzmann (SB) law (for a planar black body), with $\sigma =
\pi^2 k^2_B / (60 \hbar^3 c^2)$ and $A$ the surface area of the
spheres. The inset of the figure also shows the corresponding flux
spectra $\Phi$ of both interacting ($d=R$) and isolated spheres,
normalized by $A$ and plotted over relevant wavelengths $\lambda
\gtrsim \lambda_T$, where $\lambda_T = \hbar c / k_B T \approx
7.6\mu$m denotes the thermal wavelength corresponding to the peak of
the thermal spectrum. In both cases the BEM results (circles) are
shown to agree with the corresponding semi-analytical formulas (in the
case of isolated spheres, the flux spectrum is compared against
\eqref{spherephi}).

\section{Applications}
\label{sec:apps}

In this section, we illustrate the generality and broad applicability
of the FSC formulation by applying the BEM RWG method to obtain new
results in complex geometries. As discussed above, most calculations
of heat transfer have focused primarily on semi-infinite planar
bodies.\cite{BasuZhang09} Finite bodies only recently became
accessible with the development of sophisticated spectral
methods,\cite{Biehs08,bimonte09,Messina11,Kruger11,OteyFan11,Golyk12,Kruger12}
albeit for highly symmetric bodies with smooth shapes (e.g. spheres)
for which convenient spectral bases exist. Here we will focus instead
on geometries involving finite bodies with sharp corners (combinations
of finite plates, cylinders, and cones) that pose no challenge for the
BEM RWG method but which prove difficult to model via spectral
methods.

\begin{figure}[t!]
\includegraphics[width=1.0\columnwidth]{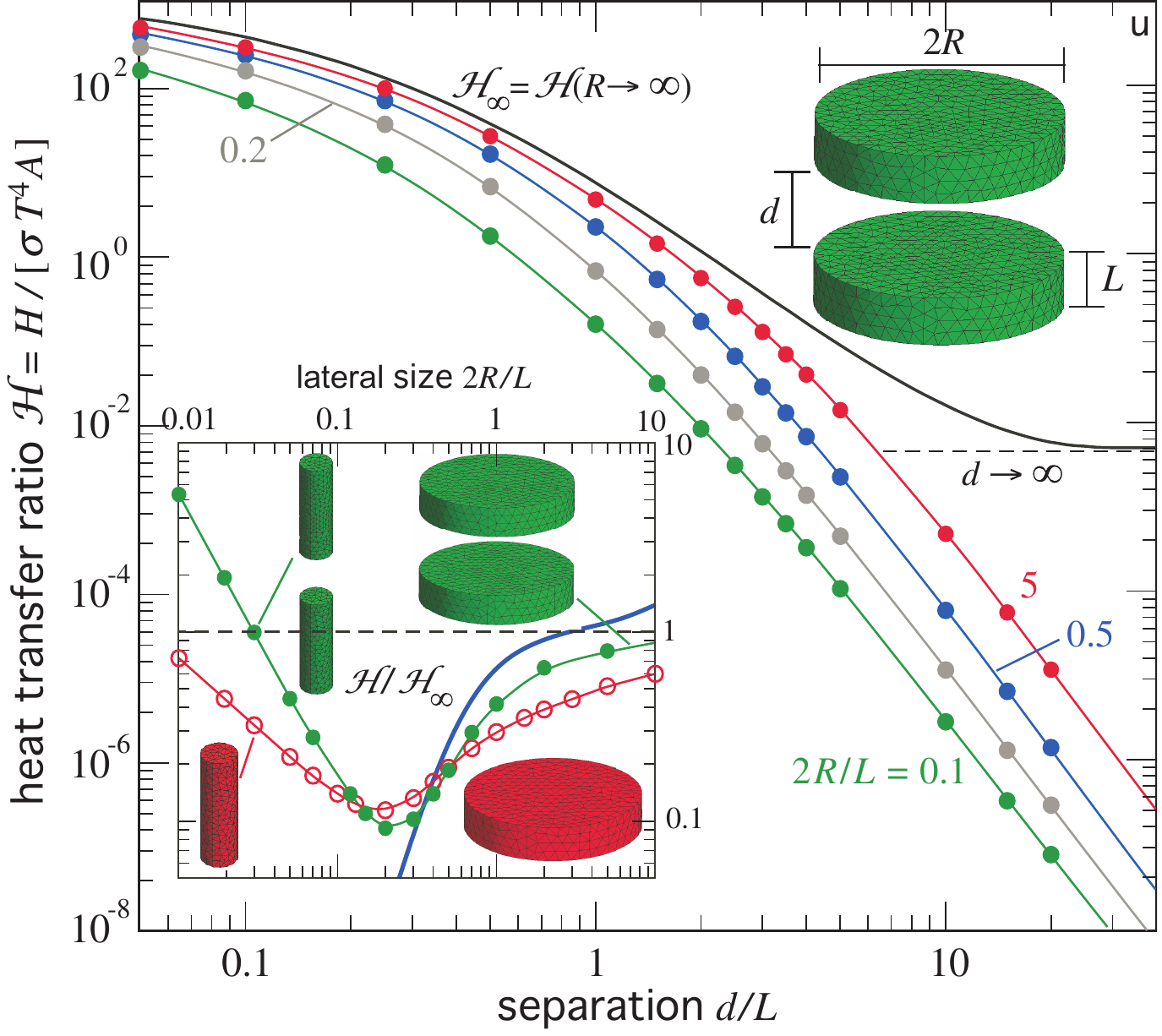}
\caption{Ratio $\h = H/\sigma T^4 A$ of the heat-transfer rate $H$
  from a finite, gold circular plate of lateral size $2R$ and
  thickness $L=0.2\mu$m held at $T=300$~K, to an identical plate held
  at zero temperature, and the SB law $\sigma T^4 A$ (where $A=\pi
  R^2$ is the ``interaction'' surface area of the plates), as a
  function of their surface--surface separation $d$. $\h$ is plotted
  for multiple aspect ratios $2R/L$ (circles). The solid black line
  corresponds to the heat-transfer ratio $\h_\infty = \h_\infty(R \to
  \infty)$ obtained upon taking the limit $R\to \infty$, which is
  computed via the semi-analytical formula in
  \citeasnoun{Ben-Abdallah09}. (Inset:) Heat-transfer rate between two
  plates at a fixed separation $d=0.1L$ (solid circles), and heat
  radiation of an isolated plate (open circles) or sphere (thick solid
  line), as a function of lateral size or diameter.}
\label{fig:fig-plates}
\end{figure}

\begin{figure}[t!]
\includegraphics[width=1.0\columnwidth]{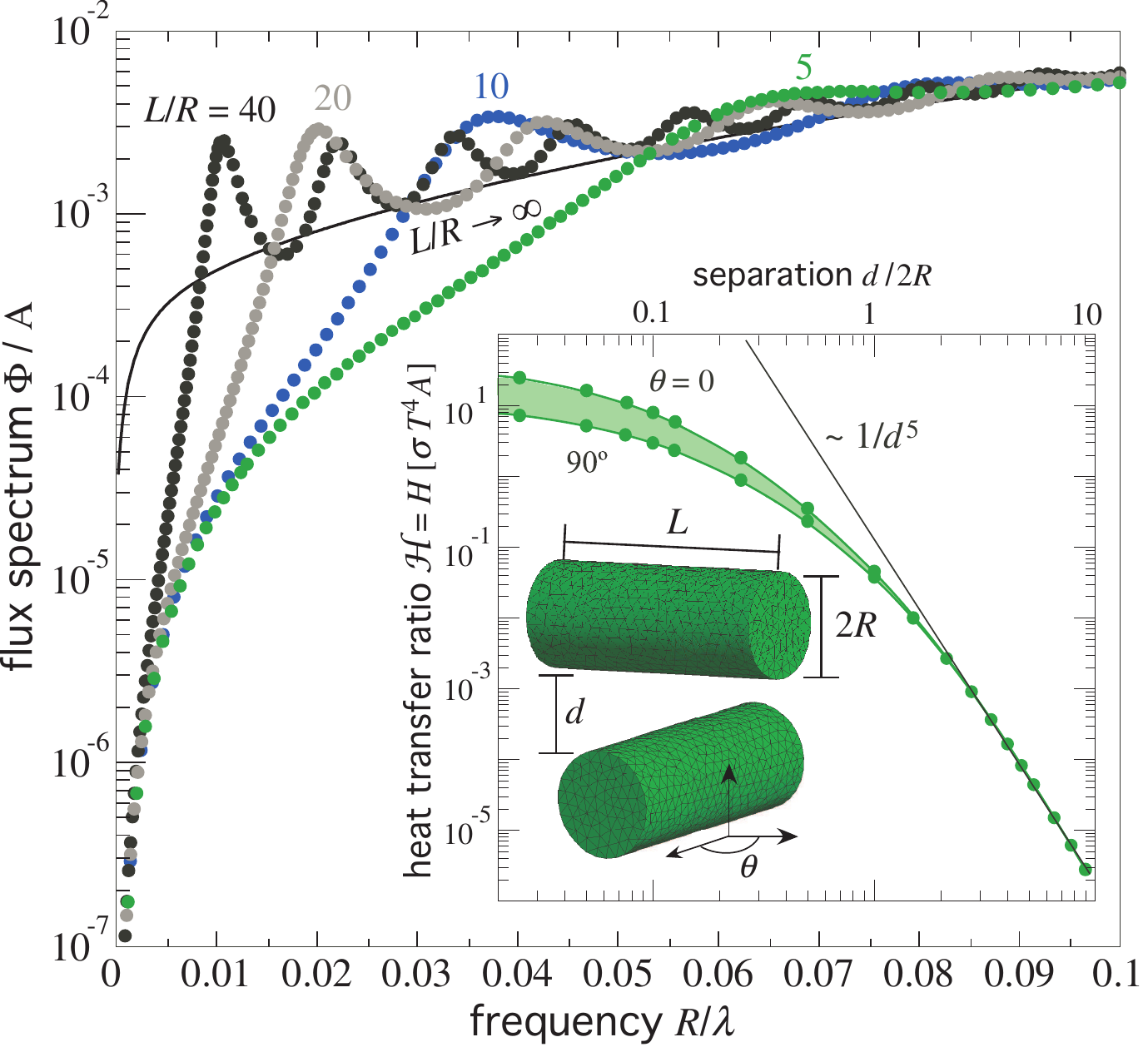}
\caption{Flux spectrum $\Phi(\omega)$ of an isolated gold cylinder of
  length $L$ and radius $R=0.2\mu$m, normalized by its corresponding
  surface area $A$, and plotted for multiple aspect ratios $L/R$
  (solid circles). The solid line shows $\Phi$ in the limit $L\to
  \infty$ of a semi-infinite cylinder, as computed by the
  semi-analytical formula of~\citeasnoun{Golyk12}. (Inset:)
  Heat-transfer ratio $\h$ of heat transfer $H$ from a
  room-temperature cylinder of aspect ratio $L/R = 5$ to an identical
  cylinder at zero temperature, and the SB law $\sigma T^4 A$, with
  $A=2\pi R(R+L)$ denoting the total surface area of each cylinder, as
  a function of their surface--surface separation $d$. $H$ is plotted
  for both parallel ($\theta=0$) and crossed ($\theta=90^\circ$)
  cylinder configurations, with the shaded region corresponding to
  intermediate $\theta$.}
\label{fig:fig-cyls}
\end{figure}

\subsection{Plates and cylinders}

To begin with, we extend the calculation of heat transfer between
planar semi-infinite plates to the case of finite plates, which
quantifies the influence of lateral size effects in that
geometry. \Figref{fig-plates} shows the ratio $\h = H/\sigma T^4 A$ of
the heat-transfer rate $H$ between finite circular plates of thickness
$L$ and lateral size $2R$ and the SB law, with one plate held at
$T=300$~K and the other held at zero temperature, as a function of
their surface--surface separation $d$. $\h$ is plotted for multiple
aspect ratios $2R/L$ (solid circles), with fixed $L=0.2\mu$m.  For
comparison, we also plot the heat-transfer ratio $\h_\infty$ for
semi-infinite ($R\to \infty$) plates of the same thickness (black
solid line), which is obtained analytically from the absorptivity of
the plates via Kirchhoff's law of thermal
radiation.\cite{Rytov89,Reif:stat} As expected, one finds that at
small $d$, near-field effects dominate and $\h \sim 1/\sqrt{d}$ for
both finite and semi-infinite plates. In contrast, at asymptotically
large $d$, the finite plates behave like dipoles and one finds that
$\h \sim 1/d^5$, whereas the semi-infinite transfer rate approaches a
constant $\h_\infty(d\to \infty) \ll 1$ independent of $d$; the rate
is significantly smaller than that of a perfect black body because
gold is highly reflective. As $R \to \infty$, the BEM results approach
$\h_\infty$ for all separations $d$, albeit at different rates, where
smaller separations converge faster than larger separations.

To quantify finite-size effects, the inset of \figref{fig-plates}
shows $\h/\h_\infty$ for isolated and interacting plates (at a single
separation $d=0.1L$) as a function of $R$. As above, in the limit of
large $R \gg \lambda_T \gg L$, such that the dominant wavelengths and
corresponding skin depths $\delta = c / \Im \sqrt{\varepsilon} \omega$
are much smaller than the lateral dimensions of the plates, $\h \to
\h_\infty$. In the case of isolated plates, the relevant lengthscales
are $\lambda_T$ and $\delta$, whereas in the case of interacting
plates, the separation $d$ also factors into the convergence rate: the
increasing contribution of (long-wavelength) near-field effects to the
heat transfer at smaller separations means that smaller separations
converge faster to the $\h_\infty$ result than larger
separations. (For the particular separation $d=0.1L$ plotted here,
near field effects are large enough to cause the convergence rate of
the interacting plates to be significantly larger than that of the
isolated plate.) At intermediate $R \lesssim L \ll \lambda_T$, the
plates no longer resemble plates but rather elongated cylinders,
leading to significant deviations in $\h$.

Compared to the heat radiation of semi-infinite cylinders ($L/R \to
\infty$ for fixed $R$), studied previously by~\citeasnoun{Golyk12},
the radiation of finite cylinders displays a number of interesting
features. (Note that $\h$ here includes radiation emitted in both the
axis-parallel, $\h_\parallel$, and -perpendicular, $\h_\perp$,
directions.) First, due to the finite value of $L$, in the limit
$R\to\infty$ the radiation of our finite cylinders is best
characterized by the radiation of thin plates with $\h_\parallel \gg
\h_\perp$. Not surprisingly, we find that $\h\to \h_\infty$ from below
as $R\to\infty$, in contrast to what is observed in the semi-infinite
case where $\h_\infty$ is approached from above.~\cite{Golyk12} Second
and most interestingly, we find that below a critical $R$, determined
by the smallest skin-depth $\delta \approx 20$nm of Au over the
relevant thermal wavelengths, the radiation normalized by surface area
\emph{increases} with decreasing $R$, leading to
non-monotonicity. Such behavior is unusual in that in this $R\lesssim
\delta$ regime, bodies most often behave like volume emitters, causing
$H$ to grow with the volumes rather than surfaces of the bodies (as
observed in the case of semi-infinite cylinders).~\cite{Golyk12}
Indeed, we find that for dielectric bodies with small and positive
$\varepsilon$, one obtains the usual volume-dependence of $H$.  In
contrast, the enhancement in \figref{fig-plates} arises because for
small $R$, the cylinders act as metallic dipole emitters, whose
radiation is increasingly dominated by $\h_\parallel$ as $R\to 0$ and
whose quasistatic (long wavelength) parallel polarizability grows with
decreasing $R$ (a consequence of the increasing anisotropy of the
cylinder and large $\Im \varepsilon$).~\cite{Hulst81,Venermo05} For
sufficiently small $R$, the heat transfer per unit area of the
uniaxial cylinders can greatly exceed that of the semi-infinite plate,
i.e. $\h \gg \h_\infty$.  (The dipole model also predicts that $\h$
will eventually vanish as $R\to 0$, but only at radii too small to be
easily calculated by BEM.  We intend to explore these phenomena more
fully in subsequent work.)

It is also interesting to study the convergence of the cylinder
radiation rate with $L$, comparing our results against the
semi-analytical results obtained in the special case of semi-infinite
($L\to\infty$) cylinders.~\cite{Golyk12} We also consider the heat
transfer between non-uniaxial (parallel) cylinders. \Figref{fig-cyls}
shows the flux spectra $\Phi$ of isolated cylinders of radius
$R=0.2\mu$m and varying lengths $L$; for comparison, we also plot the
spectrum of the semi-infinite cylinders~\cite{Golyk12} (solid lines).
As before, $\Phi$ is normalized by the surface area $A$ of each
object. (For the relevant wavelength range shown in the figure, $R$ is
several times $\delta$, which means that most of the radiation is
coming from sources near the surface of the objects.\cite{Golyk12}) We
find that for $L/R \approx 2$ (not shown), corresponding to
nearly-isotropic cylinders, $\Phi$ is only slightly larger than that
of an isolated sphere due to the small but non-negligible contribution
of volume fluctuations to $\Phi$.  As $L/R$ increases, $\Phi$
increases over all $\lambda$, and converges towards the $L \to \infty$
limit (black solid line) as $\lambda \to 0$, albeit slowly. Moreover,
$\Phi_L \gg \Phi_\infty$ at particular wavelengths, a consequence of
\emph{geometrical} resonances that are absent in the semi-infinite
case---away from these resonances, $\Phi$ clearly straddles the $L \to
\infty$ result so long as $\lambda \lesssim L$. As in the case of
finite plates, the $\Phi$ of interacting cylinders exhibits
significant enhancement at large $\lambda$ due to near-field effects,
so that $\h \to \infty$ with decreasing separation $d$. The
enhancement is evident in \figref{fig-cyls}, which shows $\h$ over a
wide range of $d$ for both parallel- ($\theta=0$) and crossed-cylinder
($\theta=90^\circ$) configurations, with one cylinder held at
$T=300$~K and the other at zero temperature (both cylinders have
aspect ratio $L/R=5$). We find once again that there are two very
distinct separation regimes of heat transfer: at large $d \gg R$, the
cylinders act like dipole emitters and $\h/\h_\infty \sim 1/d^5 \ll 1$
whereas at small $d \ll R$, flux contributions from evanescent waves
dominate and $\h/\h_\infty \sim 1/\sqrt{d} \gg 1$. Comparing the heat
transfer $H$ in the parallel and crossed-cylinder configurations, we
find that $H_\parallel / H_\perp \approx 1$ at large $d \gg R$ but
increases significantly at smaller $d \ll R$, again due to near-field
effects: in the $d\to 0$ limit, $H$ is dominated by closest
surface--surface interactions, so $H_\parallel / H_\perp \sim L/R \to
5$. As expected, $H_\parallel/H_\perp \to \infty$ as $L\to \infty$
because the increased ``interaction'' area in this limit favors the
parallel over the crossed configuration. Specifically, whereas $H$
grows linearly with $L$ in the parallel configuration, it grows
sub-linearly (and asymptotes to a finite value in the $L\to\infty$
limit) in the crossed configuration due to the diminishing
contributions of near-field and radiative flux between surface
elements in the extremities of the cylinders.



\begin{figure}[t!]
\includegraphics[width=1.0\columnwidth]{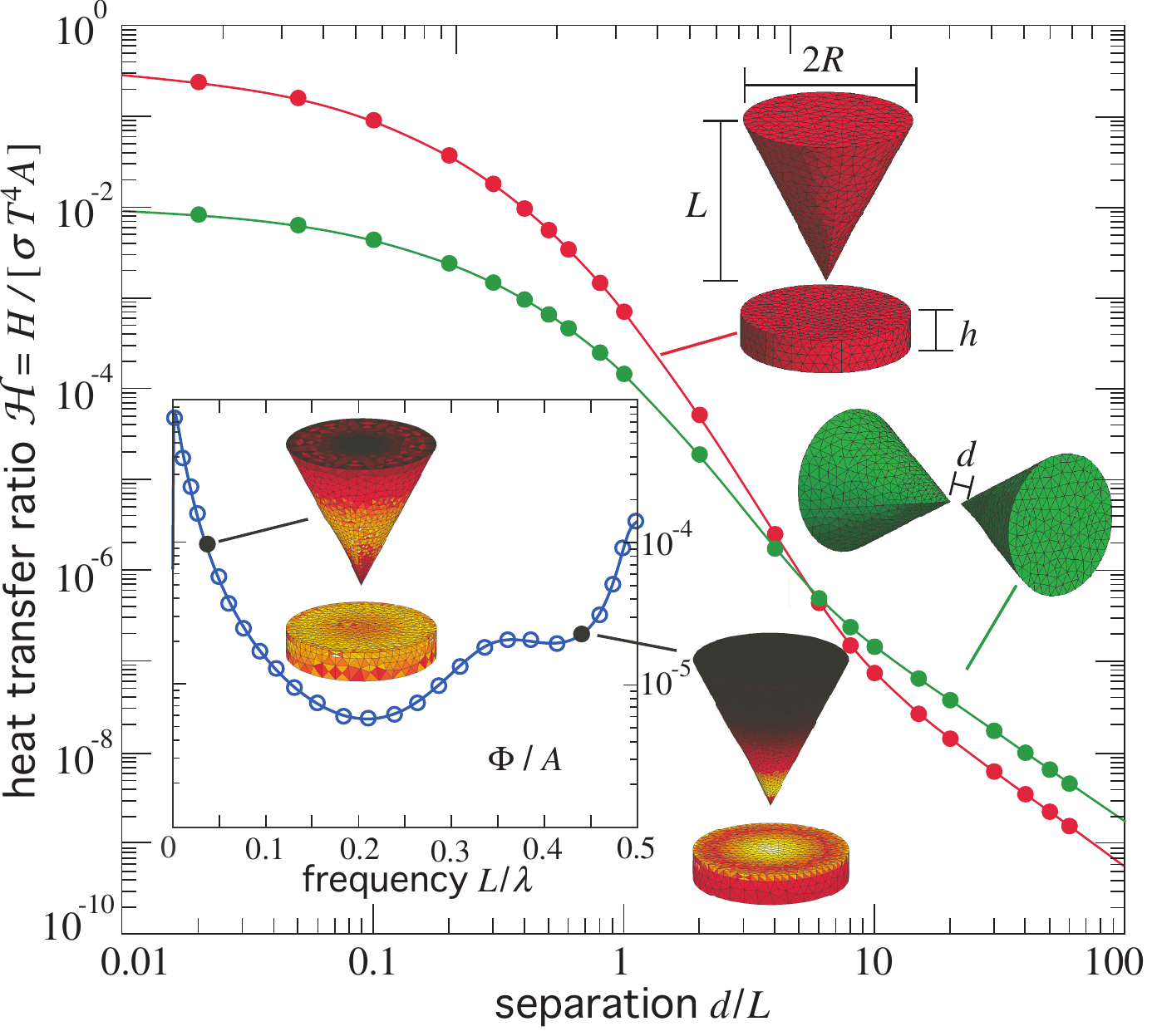}
\caption{Heat-transfer rate $H$ from a room-temperature gold cone of
  base radius $R = 1\mu$m and length $L=2R$, to either an identical
  cone (green circles) or a gold plate of radius $R$ and thickness
  $h=0.2\mu$m, held at zero temperature, as a function of their
  surface--surface separation $d$. $H$ is normalized by the
  Stefan-Boltzmann law $\sigma T^4 A$, where $A$ is the surface area
  of the cone. (Inset:) Flux spectrum $\Phi(\omega)$ of the
  cone--plate configuration at a single separation $d=0.2L$,
  normalized by the area of the cone. The two surface-contour plots
  show the distribution of flux pattern on the surfaces of the bodies
  at two wavelengths, $\lambda \approx 30 L$ and $\lambda \approx 2.2
  L$, where white/black denotes the maximum/minimum flux at the
  corresponding wavelength.}
\label{fig:fig-cones}
\end{figure}

\subsection{Cones}

Finally, motivated by recent predictions,\cite{McCauleyReid12} we
consider the heat transfer between finite cones and
plates. In~\citeasnoun{McCauleyReid12}, the cone--plate geometry (with
a semi-infinite plate) was obtained using a ``hybrid'' scattering--BEM
method~\cite{McCauleyReid12} based on the scattering--theory
formulation of~\citeasnoun{Kruger11}. (In contrast to semi-infinite
plates or spheres, the scattering-matrix of a cone cannot be easily
obtained analytically, and was instead computed numerically by
exploiting the BEM method in combination with a multipole basis of
cylindrical waves.) Here, in addition to extending these predictions
to the case of finite plates, we consider the heat-transfer rate
between two oppositely oriented cones. \Figref{fig-cones} shows the
heat-transfer rate $h$ (as in the previous section, $h=H / \sigma T^4
A$ where here $A=\pi R^2$ is the projected area of the cone) from a
cone of radius $R=0.5\mu$m and length $L=2R$ to either an identical
cone rotated by $180^\circ$ (green circles) or a plate of radius $R$
and thickness $L=0.2\mu$m (red circles), as a function of their
surface--surface separation $d$. As before, we consider gold bodies,
with one held at $300$~K while the other is held at zero temperature
Similar to~\citeasnoun{McCauleyReid12}, we find that the heat transfer
rate $H \sim \log(d)$ varies logarithmically with $d$ at short
separations $d \ll L \ll \lambda_T$, a consequence of near-field
interactions and the finite size of the cone.~\cite{McCauley:thesis}
While $h$ exhibits similar scaling with $d$ for both geometries, $h$
turns out to be almost two orders of magnitude smaller at small $d \ll
L$ in the cone--cone geometry, as would be expected from a
proximity-approximation (PA) model~\cite{Sasihithlu11}.  The situation
is reversed at large separations $d \gg \lambda_T \gg L$: beyond a
critical $d \approx 7L$, the cone--cone heat transfer becomes larger
than the cone--plate transfer. The reversal is expected on the basis
that at these separations, the two bodies act like fluctuating dipoles
oriented mainly along their largest dimension (along the axis of
symmetry for the cone and along the lateral dimension for the plate),
in which case the cone--plate interaction resembles the interaction of
two orthogonal dipoles whereas the cone--cone interaction resembles
the interaction of two parallel dipoles. Another interesting feature
of the heat-transfer in this geometry is that the spatial distribution
of pattern over the plate exhibits a local minimum directly below the
tip of the cone, a consequence of the dipolar field induced on the
cone at long wavelengths.\cite{McCauleyReid12} Here, we observe a
similar phenomenon but we find that the finite size of the plate
significantly alters the scope of the anomalous radiation pattern. In
particular, whereas \citeasnoun{McCauleyReid12} found this effect to
persist over a wide range of wavelengths (surviving even in the total
or integrated radiation pattern), we find that in the finite-plate
case it disappears much more rapidly with decreasing wavelength.

\section{Conclusion}

The FSC approach to non-equilibrium fluctuations presented here
permits the study of heat transfer between bodies of arbitrary shape,
paving the way for future exploration of heat exchange in
microstructured geometries that until now remain largely unexplored in
this context. Our formulation shares many properties with previous
scattering-matrix formulations of radiative heat transfer, e.g. our
final expressions involve traces of matrices describing scattering
unknowns, but differs in that our ``scattering unknowns'' are surface
currents defined on the surfaces of the bodies rather than incident
and outgoing waves propagating into and out of the
bodies~\cite{Narayanaswamy08:spheres,bimonte09,Messina11,Kruger11,OteyFan11,Biehs11:apl,Golyk12,Kruger12,Guerout12,Marachevsky12}. As
argued above, this choice of description has important conceptual and
numerical implications: it allows direct application of the
surface-integral equation formalism as well as the boundary-element
method. When specialized to handle high-symmetry geometries using
special functions that exploit those symmetries, our approach can be
used to obtain fast-converging semi-analytical formulas in the spirit
of previous work based on spectral
methods~\cite{Kruger11,Golyk12,Kruger12}. Moreover, it can also be
applied as a brute-force method, taking advantage of existing,
well-studied, and sophisticated BEM codes (with no modifications), to
obtain results in arbitrary/complex geometries.

While the main focus of this work was on exploring some of the ways in
which the FSC formulation can be applied to study non-equilibrium heat
transfer, we believe that analogous techniques can be used to derive
corresponding FSC approaches to other fluctuation phenomena, including
near-field fluorescence~\cite{Saidi09}, quantum noise in
lasers~\cite{Chong12}, and non-equilibrium Casimir
forces~\cite{Buhmann08,Kruger11}, an idea we plan to explore in future
work. Furthermore, although our calculations here focused on geometries
involving compact bodies, the same heat-transfer formulas derived
above apply to geometries involving infinitely extended/periodic
bodies (of importance in applications of heat transfer to
thermophotovoltaics). Modifying BEM solvers to handle periodic
structures, however, is
non-trivial~\cite{Nedelec91,Taflove00,Nicholas08,Arens10,Barnett11}
and we therefore consider that case in a subsequent publication.

Finally, although \eqref{Phi} is already well-suited for efficient
numerical implementation, its computational efficiency may be improved
by adopting a modified formulation in which the dense $G$ matrices are
replaced by certain \textit{sparse} matrices involving overlap
integrals among basis functions. In addition to reducing the
computational cost of the trace in \eqref{Phi}, this approach has the
advantage of allowing the computation of other fluctuation-induced
quantities such as non-equilibrium Casimir forces and torques.  This
alternative formulation will be discussed in a forthcoming
publication.

\section*{ACKNOWLEDGEMENTS}

This work was supported by DARPA Contract No. N66001-09-1-2070-DOD, by
the AFOSR Multidisciplinary Research Program of the University
Research Initiative (MURI) for Complex and Robust On-chip
Nanophotonics, Grant No. FA9550-09-1-0704, and by the U.S. Army
Research Office under contracts W911NF-07-D-0004 and W911NF-13-D-0001.

\appendix

\section{Eigenfunctions of the Helmholtz equation}
\label{sec:appendix-eigs}

In this section, we provide and exploit the standard Fourier and
spherical-wave eigenfunctions of the vector Helmholtz operator,
obtained by solving \eqref{Helmholtz} in planar and spherical
coordinates,\cite{Tau93,Rahi09:PRD} to obtain the coefficients $\chi$
and $\kappa$ appearing in the Green's function expansion and
orthogonality relations of \eqref{GD} and \eqreftwo{XM}{XN},
respectively.

\subsection{Fourier basis}
\label{sec:appendix-Fourier}

In planar geometries, described by normal and tangential coordinates
$z$ and $\vec{x}_\perp$, respectively, the eigenfunctions of the
Helmholtz operator, labeled by Fourier wavevectors $\vec{k}_\perp$
perpendicular to the $\hat{\vec{z}}$ axis, are given by:
\begin{align*}
  \vec{M}^\pm_{\vec{k}_\perp k_z}(z,\vec{x}_\perp) &= \phi^\pm(k_z z)
  \vec{X}_{\vec{k}_\perp}(\vec{x}_\perp),
  \\ \vec{N}^\pm_{\vec{k}_\perp k_z}(z,\vec{x}_\perp) &= \phi^\pm(k_z
  z) \left[ \mp \frac{k_z}{k} \vec{Z}_{\vec{k}_\perp}(\vec{x}_\perp) +
    \frac{|\vec{k}_\perp|}{k} e^{i\vec{k}_\perp \cdot \vec{x}}
    \vechat{z} \right],
\end{align*}
where $\phi^\pm_{\vec{k}_\perp k_z} = \frac{1}{|\vec{k}_\perp|}
e^{i\vec{k}_\perp \cdot \vec{x}_\perp \pm i k_z z}$, $k_z = \sqrt{k^2
  - |\vec{k}_\perp|^2}$, and where the tangential fields
$\vec{X}_{\vec{k}_\perp}$ and $\vec{Z}_{\vec{k}_\perp} = \vechat{z}
\times \vec{X}_{\vec{k}_\perp}$ are:
\begin{align}
  \vec{X}_{\vec{k}_\perp}(\vec{x}_\perp) &= \frac{i}{|\vec{k}_\perp|}
  \left(\hat{\vec{z}} \times \vec{k}_\perp \right) e^{i\vec{k}_\perp
    \cdot \vec{x}_\perp} \\ \vec{Z}_{\vec{k}_\perp}(\vec{x}_\perp) &=
  \frac{i \vec{k}_\perp}{|\vec{k}_\perp|} e^{i\vec{k}_\perp \cdot
    \vec{x}_\perp}
\end{align}
The precise form of the Fourier functions $\phi^\pm = e^{\pm ik_z z}$
depends on whether one desires a solution involving outgoing ($+$) or
incoming ($-$) fields, or equivalently, fields propagating away or
toward the origin.  The corresponding $\chi$ and $\kappa$ coefficients
appearing in the Green's function expansion and orthogonality
relations are given by:
\begin{align}
  \kappa^{\pm}_{\vec{k}_\perp, k_z, E}(z) &= \phi^{\pm}(k_z z)
  \\ \kappa^{\pm}_{\vec{k}_\perp, k_z, M}(z) &= \mp \gamma
  \phi^{\pm}(k_z z) \\ \chi_{\vec{k}_\perp, k_z} &= \frac{i}{2 k_z}
\end{align}
with $\gamma \equiv k_z / k = \sqrt{1 - |\vec{k}_\perp|^2/(\varepsilon
  \mu \omega^2)}$.

\subsection{Spherical multipole basis}
\label{sec:appendix-mult}

In spherically symmetric geometries, described by normal and
tangential coordinates $r$ and $\{\theta,\phi\}$, respectively, the
eigenfunctions of the Helmholtz operator, labeled by angular-momentum
quantum numbers $\ell$ and $m$, are given by:
\begin{align*}
  \vec{M}^{\pm}_{\ell m}(r,\theta,\phi) &= R^{\pm}_\ell(kr)
  \vec{X}_{\ell m}(\theta,\phi), \\ \vec{N}^{\pm}_{\ell m}(r,
  \theta, \phi) &= \breve{R}^{\pm}_\ell(kr) \vec{Z}_{\ell m} +
  \frac{\ell(\ell+1)}{r} R^{\pm}_{\ell}(kr) Y_{lm}(\theta,\phi)
  \vechat{r},
\end{align*}
where $R^{\pm}_\ell$ and $Y_{\ell m}$ denote spherical Hankel
functions and spherical harmonics,\cite{Jackson98} respectively, and
where the tangential fields $\vec{X}_{\ell m} =
-\frac{1}{\sqrt{\ell(\ell+1)}}(\vechat{r} \times \nabla) Y_{\ell m}$
and $\vec{Z}_{\ell m} = \vechat{r} \times \vec{X}_{\ell m}$ are:
\begin{align*}
  \vec{X}_{\ell m}(\theta, \phi) &= \frac{1}{\sqrt{l(l+1)}} \Big[
    \frac{im}{\sin\theta} Y_{\ell m} \vechat{\theta} -\frac{\partial
      Y_{\ell m}}{\partial \theta} \vechat{\varphi} \Big]
  \\ \vec{Z}_{\ell m}(\theta,\phi) &= \frac{1}{\sqrt{l(l+1)}} \Big[
    \frac{\partial Y_{\ell m}}{\partial \theta} \vechat{\theta} +
    \frac{im}{\sin\theta} Y_{\ell m} \vechat{\varphi} \Big].
\end{align*}
Above, we defined $\breve{f}(z) \equiv (1/z + d/dz) f$ for
brevity. The precise form of the spherical Bessel radial function
\begin{equation*}
  R^{\pm}_\ell = 
  \begin{cases}
    h^{(1)}_\ell & + \\
    j_\ell & -
  \end{cases}
\end{equation*}
depends on whether one desires a solution corresponding to outgoing
($+$) or incoming ($-$) waves toward the origin, or equivalently, a
solution that is well-behaved at the origin or at infinity. The $\chi$
and $\kappa$ coefficients appearing in the Green's function expansions
and orthogonality relations are given by:
\begin{align}
  \kappa^{\pm}_{\ell, m, E}(r) &= r^2 R^{\pm}_\ell(kr) \\ 
  \kappa^{\pm}_{\ell, m, M}(r) &= i r^2 \breve{R}^{\pm}_\ell(kr) \\ 
  \chi_{lm} &= i k
\end{align}
 
\section{Equivalence principle}
\label{sec:appendix-equiv}

In this section, we provide a compact derivation and review of the
equivalence principle of classical electromagnetism (closely related
to Huygens' principle\cite{Merewether80}), which expresses scattered
waves in terms of fictitious equivalent currents in a homogeneous
medium replacing the scatterer.\cite{Harrington89} The equivalence
principle is usually derived in a somewhat cumbersome way from a
Green's-function approach,\cite{Harrington89,Chen89} but a much
shorter proof can be derived from the differential form of Maxwell's
equation. Understanding this result is central to our FSC formulation
of heat transfer.

As before, we restrict ourselves to linear media, for which Maxwell's
equations can be written as:
\begin{equation}
  \underbrace{\mat{ & \nabla \times \\ -\nabla \times &}}_M
  \underbrace{\mat{\vec{E} \\ \vec{H}}}_{\f} =
    \frac{\partial}{\partial t} \left[ \f + \chi \star \f\right] +
    \underbrace{\mat{\vec{J} \\ \vec{K}}}_\x,
\end{equation}
with $\chi {}\star$ denoting convolution with a $6\times 6$
susceptibility tensor
\begin{equation*}
  \chi = \mat{\varepsilon-1 & \\ & \mu-1}.
\end{equation*}

Consider an arbitrary incident wave $\f$ which solves the
\emph{source-free} Maxwell's equations in some $\chi$ medium with
\emph{no} current sources: $M\f = \frac{\partial}{\partial t} [ \f +
  \chi \star \f]$. The equivalence principle states that, given any
arbitrary but \emph{finite} domain $V$, one can always choose an
\emph{equivalent} surface current $\x$ that generates the same
incident field $\f$ in $V$. To show that such a surface current
exists, define the field
\begin{equation}
  \tilde{\f} =
  \begin{cases}
    \f & \in V, \\
    0 & \mathrm{elsewhere}
  \end{cases}
\end{equation}
It follows that $\tilde{\f}$ satisfies the source-free Maxwell's
equations in \emph{both} the interior and exterior regions---the only
question is what happens at the interface $\so$. In particular,
the discontinuity of $\tilde{\f}$ at $\so$ produces a surface delta
function $\delta_{\so}$ in the spatial derivative $M\tilde{\f}$, and so in
order to satisfy Maxwell's equations with this $\tilde{\f}$ one must
introduce a matching delta function, a \emph{surface current} $\x$, on
the right-hand side. [Here, $\delta_{\so}$ is the distribution such that
  $\iiint \delta_{\so} \f(\vec{x}) = \oiint_{\so} \f(\vec{x})$ for any
  continuous test function $\f$.]  Specifically, letting $\vec{n}$ be
the unit inward-normal vector~\footnote{For simplicity, we assume a
  differentiable surface $\so$ so that its normal $\vec{n}$ is well
  defined, but the case of surfaces with corners (e.g. cubical
  domains) follows as a limiting case.}, only the normal derivative
$\vec{n} \cdot \nabla$ contains a delta function (whose amplitude is
the magnitude of the discontinuity), which implies a surface current:
\begin{equation}
  \x = (\Theta \f) \delta_{\so} = \mat{\vec{n} \times \vec{H}
    \\ -\vec{n}\times \vec{E}} \delta_{\so},
\end{equation}
where $\Theta$ is the real-symmetric unitary
$6\times6$ matrix:
\begin{equation}
  \Theta = \mat{& \vec{n}\times \\ -\vec{n}\times & } = \Theta^{-1} = \Theta^\T = \Theta^*.
\end{equation}
That is, there is a surface \emph{electric} current given by the
surface-tangential components $\vec{n}\times \vec{H}$ of the incident
\emph{magnetic} field, and a surface \emph{magnetic} current given by
the components $-\vec{n} \times \vec{E}$ of the incident
\emph{electric} field. These are the equivalent currents of the
principle of equivalence (derived traditionally from a Green's
function approach~\cite{Harrington89,Chen89} and from which Huygens's
principle is derived~\cite{Merewether80}).

\subsection{Application to surface integral equations}

The equivalence principle is of fundamental importance to SIE
formulations of EM scattering.  Consider two regions 0 and 1,
described by volumes $V^\o{0}$ and $V^\o{1}$ and susceptibilities
$\chi^\o{0}$ and $\chi^\o{1}$, respectively, separated by an interface
$\so^\o{1}$. As before, one can express the total fields $\f^\o{r} =
\f^\op{r} + \f^\om{r}$ in each region $r$, in terms of incident
$\f^\op{r}$ and scattered $\f^\om{r}$ fields. The principle of
equivalence describes an equivalent, fictitious problem, involving
fields
\begin{equation*}
  \tilde{\f} =
  \begin{cases}
    \f^\o{0} & \in V^\o{0}, \\
    0 & \mathrm{elsewhere}
  \end{cases}
\end{equation*}
and surface currents $\x = \Theta \f^0 = \Theta \f^1$ at the
$\so^\o{1}$ interface, where the second equality follows from
continuity of the tangential fields. Since $\tilde{\f} = 0$ in
$V^\o{1}$, it follows that one can replace $\chi^\o{1}$ with any other
local medium and yet $\tilde{\f}$ still satisfies Maxwell's
equations. In particular, replacing $\chi^\o{1}$ with $\chi^\o{0}$
implies that one can write the scattered field $\f^\om{0} = \G^\o{0}
\star \xi$ in $V^\o{0}$ as the field produced by the same fictitious
surface currents $\x$ in an \emph{infinite} medium 0, with $\G^\o{0}$
denoting the \emph{homogeneous-medium} Green's function of the
infinite medium.

A similar argument applies if one is interested in the field in medium
1, except that the sign of the fictitious currents is reversed to
$-\x$ in order to account for the direction of the discontinuity in
going from 1 to 0 in this case. In particular, one can write the
scattered field $\f^\om{1} = - \G^\o{1} \star\x$ in $V^\o{1}$ as the
field produced by a fictitious surface current $-\x$ in an infinite
medium 1.

\section{Reciprocity and definiteness}

In this section, we present a brief review of the reciprocity
relations and definiteness (positivity) properties of the DGF, $\G$,
connecting surface currents $\x$ to fields $\f = \G \star \x$, in
dissipative media, and explain how these relate to corresponding
properties of the SIE matrices above (crucial to our derivation of
heat transfer in \secref{FSC-deriv}). Although for our purposes we
need only prove reciprocity and definiteness of the \emph{homogeneous}
Green's function (trivial to show in that case since the homogeneous
DGF is known analytically), here we consider the more general case of
inhomogeneous media. Reciprocity is well
known~\cite{Lifshitz80,Altman91,Karlsson92,Born99,Potton04}, and
positivity follows from general physical principles (currents always
do nonnegative work in passive
materials~\cite{Landau:EM,Reif:stat,Karlsson92,Jackson98}), but our
goal here is to derive them using the same language employed in our
derivations above. More specifically, we explain the source of the
sign-flip matrices $\mathcal{S}$ and $S$, which often go unmentioned
because many authors consider only $3\times 3$ Green's functions
(relating currents to fields of the same type).

\subsection{Green's functions}
\label{sec:appendix-G-reciprocity}

It is actually easier to derive the reciprocity and definiteness
properties of $\G$ from the properties of $L = (\G \star)^{-1}$, the
Maxwell operator that connects fields $\f$ to currents $\x = L\f$,
because $L$ is a partial-differential operator that can be written
down explicitly starting from the (frequency-domain) Maxwell equations
$\nabla \times \vec{E} = i\omega\mu \vec{H} - \vec{M}$, $\nabla \times
\vec{H} = -i\omega \varepsilon \vec{E} + \vec{J}$, in terms of the
permittivity $\varepsilon(\vec{x},\omega)$ and permeability
$\mu(\vec{x},\omega)$ tensors and electric $\vec{J}$ and magnetic
$\vec{M}$ currents. Specifically, the Maxwell operator
\begin{equation}
  L = \mat{i\omega\varepsilon & \nabla \times \\ -\nabla \times &
    i\omega\mu}
\end{equation}
is neither complex-symmetric, Hermitian, anti-symmetric, nor
anti-Hermitian in general. Using our previous definition of the inner
product:
\begin{align*}
  \langle \phi, \phi' \rangle &= \int \phi^* \phi' = \left\langle
  \mat{\vec{E} \\ \vec{H}}, \mat{\vec{E}' \\ \vec{H}'} \right\rangle \\
  &= \int \vec{E}^* \cdot \vec{E}' + \vec{H}^* \cdot \vec{H}',
\end{align*}
it follows that the \emph{off}-diagonal part of $L$ is anti-Hermitian:
\begin{widetext}
\begin{align*}
  \left\langle \mat{\vec{E} \\ \vec{H}}, \mat{& \nabla \times
    \\ -\nabla \times &} \mat{\vec{E}' \\ \vec{H}'} \right\rangle 
  &= \int \vec{E}^* \cdot \nabla \times \vec{H}' -
  \vec{H}^* \cdot \nabla \times \vec{E}'  \\
  &= \int (\nabla \times \vec{E}^*) \cdot \vec{H}' - (\nabla \times \vec{H})^* \cdot \vec{E}' 
  = \left\langle -\mat{& \nabla  \times \\ -\nabla \times &} \mat{\vec{E} \\ \vec{H}}, \mat{\vec{E}' \\ \vec{H}'} \right\rangle,
\end{align*}
\end{widetext}
where we have used the self-adjointness of $\nabla \times$ and assumed
boundary conditions such that the $\oiint \vec{E}^* \times \vec{H}' +
\vec{E}' \times \vec{H}^*$ boundary terms at infinity (from the
integration by parts) vanish. This is commonly attained by assuming
loss in the materials so that the fields decay exponentially at
infinity (assuming localized sources), or by imposing
outgoing-radiation boundary conditions on $\G \star {}$ at
infinity~\cite{Jackson98}.

Instead, reciprocity relations are normally derived for the
\emph{unconjugated} inner product: 
\begin{align}
  (\phi, \phi') &= \int \phi^\T \phi' = \left(\mat{\vec{E} \\ \vec{H}},
  \mat{\vec{E}' \\ \vec{H}'}\right) \nonumber \\
  &= \int \vec{E}^\T \cdot \vec{E}' +
  \vec{H}^\T \cdot \vec{H}',
\end{align}
under which the off-diagonal terms in $L$ are still anti-symmetric
while the diagonal terms are complex-symmetric, assuming
\emph{reciprocal materials}: $\varepsilon^\T = \varepsilon$ and
$\mu^\T = \mu$ (usually the case except for magneto-optical and other
more exotic materials~\cite{Lifshitz80,Lindell94,King63}). Here, the
transpose $L^\T$ of the operator $L$ means the adjoint of $L$ under
the unconjugated inner product $(\phi, L \phi') = (L^\T \phi,
\phi')$. In order to make $L$ fully symmetric, it suffices to flip the
sign of the magnetic components $\vec{H} \to -\vec{H}$, an operation
that can be expressed as a (real, self-adjoint, unitary) sign-flip
matrix:
\begin{equation}
  \S = \mat{I & \\ & -I} = \S^{-1} = \S^\T = \S^*
\end{equation}
That is, $L\S$ is complex-symmetric: $(L\S)^\T = \S L^\T = L \S$, or
equivalently, $L^\T = \S L \S = \S L \S^{-1}$. It follows that,
\begin{equation}
  (\G \star)^\T = (L^{-1})^\T = (L^\T)^{-1} = \S (\G \star) \S
\end{equation}
Alternatively,
\begin{equation}
  (\phi, \G \star \phi') = \int\int d^3\vec{x} d^3\vec{y}
  \phi^\T(\vec{x}) \G(\vec{x},\vec{y}) \phi'(\vec{y}),
\end{equation}
so by inspection $(\G(\vec{x},\vec{y}) \star)^\T =
\G(\vec{y},\vec{x})^\T \star$: transposing $\G \star$ corresponds to
interchanging sources and fields. Thus, we obtain the reciprocity
relation:
\begin{equation}
  \G^\T \star  = \S (\G \star) \S,
\end{equation}
i.e. one can interchange sources and fields if one flips the sign of
both magnetic currents and magnetic fields.

We also expect the operators $L$ and $\G \star$ to be
negative-semidefinite on physical grounds, since $-\frac{1}{2} \langle
\f, L\f\rangle = -\frac{1}{2} \langle \f, \x \rangle = -\frac{1}{2}
\langle \G\star \x, \x\rangle$ is exactly the time-average power
$-\frac{1}{2} \int \vec{E}^*\cdot \vec{J} + \vec{H}^* \cdot \vec{M}$
expended \emph{by} the currents, which must be $\ge 0$ in passive
materials.~\cite{Landau:EM} Indeed, one can show this directly, since
the anti-Hermitian property of the off-diagonal part of $L$ means that
\begin{equation*}
  \sym L = \omega \mat{-\Im \varepsilon & \\ & -\Im \mu}
\end{equation*}
for isotropic materials. But both $\omega \Im \varepsilon$ and $\omega
\Im \mu$ are $\ge 0$ for passive materials (no
gain).\cite{Jackson98,Landau:EM} Thus, it follows that $L$ is
negative-semidefinite, and so is $L^{-1} = \G\star{}$.

\subsection{SIE matrices}
\label{sec:appendix-BEM-reciprocity}

The SIE matrices $M=W^{-1}$ are formed from a sum $\M$ of Green's
function operators $\G^\o{r} \star$ in homogeneous regions $r$,
expanded in a (real vector-valued) basis $\b_n$ by a Galerkin method,
so that $M_{mn} = \langle \b_m, \M \b_n\rangle = (\b_m, \M \b_n)$. For
any Galerkin method, it is easy to show that if $\M$ is self-adjoint
or complex-symmetric, then $M$ has the same properties. Similarly, any
definiteness of $\M$ carries over to $M$. From the previous section,
since $\G^\o{r}$ is negative-semidefinite in any passive medium, it
follows that any sum $\M$ of $\G^\o{r} \star$ is also
negative-semidefinite, and hence $\M$ is negative-semidefinite ($\sym
M$ is Hermitian negative-semidefinite).

As above, reciprocity requires some sign flips: $\M^\T = S \M S$, so
$(M^\T)_{mn} = M_{nm} = (\b_n,\M\b_m) = (\b_m,\M^\T\b_n) = (\b_m,S\M
S\b_n) = (S\b_m,\M S\b_n)$. Furthermore, suppose that we use separate
basis functions $\b_n^H$ for magnetic currents and $\b_n^E$ for
electric currents, as is typically the case in BEM (e.g. for an RWG
basis~\cite{Rao82,Rao99}), so that $S \b_n^E = +\b_n^E$ and $S\b_n^H =
-\b_n^H$. That is, we write currents as $\x = \sum_n x_n \b_n = \sum
x_n^E \b_n^E + x_n^H \b_n^H$, so that $S\x = \sum x_n^E \b_n^E - x_n^H
\b_n^H$ corresponds to a linear transformation $S$ on $x$ that flips
the sign of the $x_n^H$ components. It follows that
\begin{equation}
  M^\T = S M S.
\end{equation}


\end{document}